\documentclass[aip,jcp,preprint]{revtex4-1}

\usepackage{amsfonts}
\usepackage{graphicx}
\usepackage{amsmath}
\usepackage{gensymb}
\usepackage{color}

\begin{document}

\title{
\begin{center}
Probing Light-Induced Conical Intersections by Monitoring Multidimensional Polaritonic Surfaces
\end{center}
}

\author{Csaba F\'abri}
\email{ficsaba@staff.elte.hu}
\affiliation{MTA-ELTE Complex Chemical Systems Research Group, P.O. Box 32, H-1518 Budapest 112, Hungary}
\affiliation{Department of Theoretical Physics, University of Debrecen, PO Box 400, H-4002 Debrecen, Hungary}

\author{G\'abor J. Hal\'asz}
\affiliation{Department of Information Technology, University of Debrecen, P.O. Box 400, H-4002 Debrecen, Hungary}

\author{\'Agnes Vib\'ok}
\email{vibok@phys.unideb.hu}
\affiliation{Department of Theoretical Physics, University of Debrecen, PO Box 400, H-4002 Debrecen, Hungary}
\affiliation{ELI-ALPS, ELI-HU Non-Profit Ltd, H-6720 Szeged, Dugonics t\'er 13, Hungary}


\begin{abstract}
The interaction of a molecule with the quantized electromagnetic field of a nano-cavity gives rise to light-induced conical intersections
between polaritonic potential energy surfaces. We demonstrate for a realistic model of a polyatomic molecule that
the time-resolved ultrafast radiative emission of the cavity enables to follow both nuclear wavepacket dynamics on and nonadiabatic population
transfer between polaritonic surfaces without applying a probe pulse. The latter provides an unambiguous (and in principle experimentally accessible)
dynamical fingerprint of light-induced conical intersections.
\end{abstract}

\maketitle

Polaritonic chemistry is a rapidly emerging field
providing new means to manipulate and control light-matter
interaction.\cite{Garcia-Vidal2021,Ribeiro20186325,Herrera2020}
Coupling a quantized mode of the radiation field of an optical
or plasmonic nano-cavity to a molecular electronic or vibrational transition
creates so-called polaritonic states which are hybrid-light matter states
carrying both photonic and excitonic properties.
Polaritonic states can be viewed as the
quantum analogues of the well-known semiclassical molecular dressed
states.\cite{Bandrauk3}

Several experimental \cite{Hutchison20121592,Ebbesen20162403,cavity_Zhong_AngChem_2016,Chikkaraddy2016127}
and theoretical \cite{Galego2015,Kowalewski20162050,Flick20173026,Herrera2017,Feist2018205,Szidarovszky20186215,Vendrell2018,Csehi2019a,Fregoni2018,Mandal20195519,Perez2019,Triana2019,Ulusoy20198832,Reitz2019,Ojambati2019,Davidsson2020234304,Felicetti20208810,Fabri2020234302,Fabri2021a,Gu20201290,Silva2020,Yuen-Zhou2020,20FrCoPe,Cederbaum2021,Cederbaum2021a}
studies have demonstrated that strong light-matter coupling can lead
to the enhancement or suppression of photophysical and photochemical
processes, control charge and energy transfer, \cite{Du2018,Semenov2019,Schfer2019,Mandal2020}
accelerate singlet fission, \cite{Martinez2018} provide direct information
on quantum dynamics \cite{Silva2020} and give rise to light-induced nonadiabatic
effects.\cite{Kowalewski20162050,Feist2018205,Vendrell2018,Szidarovszky20186215,Csehi2019a,Csehi2019,Mandal20195519,Gu20201290,Gu2020a,D1CP00943E,Triana2021}
In the latter case the radiation field strongly mixes the molecular
vibrational, rotational and electronic degrees of freedom, thus creating
light-induced avoided crossings (LIACs) or light-induced conical intersections
(LICIs) between polaritonic potential energy surfaces (PESs), leading to the breakdown
of the Born--Oppenheimer (BO) adiabatic approximation.\cite{Born1927}
Similarly to natural nonadiabatic phenomena inherently present in polyatomic
molecules,\cite{Domcke2004,Cederbaum2005,Baer2006}
cavity-induced nonadiabaticity can also significantly modify the spectroscopic
\cite{Szidarovszky20186215,Fabri2020234302,Fabri2021a} and dynamical
\cite{Vendrell2018,Csehi2019a} properties of molecules. We note
that nonadiabatic effects can be induced by classical laser light as
well, which has been studied extensively over
the last decade.\cite{Sindelka2011,Corrales2014785,Halasz2015348,Natan2016,Csehi2016479,Kubel2020}

The present work focuses on monitoring nonadiabatic wavepacket dynamics
through LICIs between multidimensional polaritonic PESs.
A molecule is placed into a lossy plasmonic nano-cavity and the ultrafast photon emission
from the cavity is examined as a function of time.
One can expect significant emission in those regions of the nuclear configuration space
where the character of a given polaritonic PES is primarily photonic, therefore, the emission depends on
nuclear positions and it can be utilized to follow wavepacket motion on polaritonic PESs.
This method has been recently suggested by Feist and co-workers and successfully
applied to track the essentially BO nuclear wavepacket dynamics on polaritonic PESs.\cite{Silva2020}
The method presented in Ref. \onlinecite{Silva2020} provides a novel and ingenious route to image
nuclear wavepacket dynamics without employing probe pulses.
Our present study goes beyond the results reported in Ref. \onlinecite{Silva2020} by demonstrating
that measuring the time-dependent radiative emission also offers the possibility to monitor nonadiabatic population transfer
between polaritonic PESs through the LICI, which can be viewed as indirect probing of light-induced nonadiabaticity.
The four-atomic H$_{2}$CO (formaldehyde) molecule serves as the system of interest in the present study.
We describe H$_{2}$CO with a realistic two-dimensional vibrational model which allows the formation of LICIs
between polaritonic PESs and correctly accounts for light-induced nonadiabatic effects.
In earlier work we have shown for a single H$_{2}$CO molecule coupled to a cavity mode that the BO approximation
applied to polaritonic PESs strikingly fails due to light-induced nonadiabaticity regardless
of the number of nuclear degrees of freedom applied.\cite{Fabri2021a}

A molecule coupled to a single cavity mode can be described by the Hamiltonian \cite{04CoDuGr}
\begin{equation}
	\hat{H}_\textrm{cm} = \hat{H}_0 + \hbar \omega_\textrm{c} \hat{a}^\dag \hat{a} - g \hat{\vec{\mu}} \vec{e} (\hat{a}^\dag + \hat{a})
   \label{eq:Hcm}
\end{equation}
where $\hat{H}_0$ is the Hamiltonian of the isolated molecule, $\omega_\textrm{c}$ denotes the angular frequency of the cavity mode,
$\hat{a}^\dag$ and $\hat{a}$ are creation and annihilation operators associated with the cavity mode, $g$ is the coupling strength parameter,
$\hat{\vec{\mu}}$ corresponds to the electric dipole moment operator of the molecule and $\vec{e}$ refers to the polarization vector.
Note that the quadratic dipole self-energy term\cite{20ScRuRo,20MaMoHu,20TaMaZh,Triana2021} is neglected in Eq. \eqref{eq:Hcm}
as it is expected to add small shifts to the ground- and excited-state PESs of the molecule
in the two-dimensional vibrational model used in this work.

Considering two electronic states, labeled with X and A, the Hamiltonian of Eq. \eqref{eq:Hcm} takes the form
\begin{equation}
    \resizebox{0.9\textwidth}{!}{$\hat{H}_\textrm{cm}  = 
         \begin{bmatrix}
            \hat{T} + V_\textrm{X} & 0 & 0 & W_1 & 0 & 0 & \dots \\
            0 & \hat{T} + V_\textrm{A} & W_1 & 0 & 0 & 0 & \dots \\
            0 & W_1 & \hat{T} + V_\textrm{X} + \hbar\omega_\textrm{c} & 0 & 0 & W_2 & \dots \\
            W_1 & 0 & 0 &\hat{T} + V_\textrm{A} + \hbar\omega_\textrm{c} & W_2 & 0 & \dots \\
            0 & 0 & 0 & W_2 &\hat{T} + V_\textrm{X} + 2\hbar\omega_\textrm{c} & 0 & \dots \\
            0 & 0 & W_2 & 0 & 0 &\hat{T} + V_\textrm{A} + 2\hbar\omega_\textrm{c} & \dots \\
            \vdots & \vdots & \vdots & \vdots & \vdots & \vdots & \ddots 
        \end{bmatrix}$}
    \label{eq:cavity_H}
\end{equation}
in the direct product basis of electronic states $| \alpha \rangle$ ($\alpha = \textrm{X}, \textrm{A}$)
and Fock states $| n \rangle$ ($n=0,1,2,\dots$) of the cavity mode.
In Eq. \eqref{eq:cavity_H} $\hat{T}$ corresponds to the kinetic energy operator, $V_\textrm{X}$ and $V_\textrm{A}$
denote the ground-state and excited-state PESs and $W_n = -g \sqrt{n} \vec{d} \vec{e}$ with
$\vec{d}$ being the transition dipole moment vector. Note that terms containing the permanent dipole moments
of electronic states X and A are omitted in Eq. \eqref{eq:cavity_H}.
The polaritonic PESs can be obtained as eigenvalues of
the potential energy part of the Hamiltonian of Eq. \eqref{eq:cavity_H} at each nuclear configuration.
Of particular importance for this study is the so-called singly-excited subspace
(molecule in its ground electronic state with one photon and molecule in
the excited electronic state with zero photon). The singly-excited subspace
accommodates the lower (LP) and upper (UP) polaritonic states which can be approximately described as
superpositions of the states $| \textrm{X} \rangle |1\rangle$ and $| \textrm{A} \rangle |0\rangle$,
thus carrying both photonic and excitonic characters.

The Hamiltonian of Eq. \eqref{eq:Hcm} assumes an infinite lifetime for field excitations.
However, finite photon lifetimes can often not be neglected, which is typical for lossy plasmonic
nano-cavities.\cite{TorresSnchez2021}
In what follows, we resort to the computational protocol used in Ref. \onlinecite{Silva2020}.
In order to account for finite photon lifetimes and inherent decoherence effects,\cite{20Manzano}
we describe the quantum dynamics of the system with the density
operator $\hat{\rho}$ obeying the Lindblad master equation \cite{20Manzano,Davidsson2020234304,Silva2020}
\begin{equation}
\frac{\partial \hat{\rho}}{\partial t} =
		-\frac{\textrm{i}}{\hbar} [\hat{H},\hat{\rho}] + \gamma_\textrm{c} \hat{a} \hat{\rho} \hat{a}^\dag -
		 \frac{\gamma_\textrm{c}}{2} ( \hat{\rho} \hat{N} + \hat{N} \hat{\rho} )
	\label{eq:Lindblad}
\end{equation}
where $\hat{N} = \hat{a}^\dag \hat{a}$ is the photon number operator and $\gamma_\textrm{c}$ denotes the cavity decay rate.
The last two terms of Eq. \eqref{eq:Lindblad} containing $\gamma_\textrm{c}$ describe the incoherent decay of the cavity mode.
We have applied the values $\gamma_\textrm{c} = 10^{-4} ~ \textrm{au}$ and $\gamma_\textrm{c} = 5 \cdot 10^{-5} ~ \textrm{au}$
which are equivalent to a lifetime of $1 / \gamma_\textrm{c} =  241.9 ~ \textrm{fs}$ and
$1 / \gamma_\textrm{c} =  483.8 ~ \textrm{fs}$, respectively.
Following Ref. \onlinecite{Silva2020}, we choose to pump the cavity mode with a laser pulse and employ the Hamiltonian
\begin{equation}
	\hat{H} = \hat{H}_\textrm{cm} - \mu_\textrm{c} E(t) (\hat{a}^\dag+\hat{a})
  \label{eq:fullH}
\end{equation}
with $\mu_\textrm{c} = 1.0 ~ \textrm{au}$ (effective dipole moment of the cavity mode) and 
$E(t) = E_0 \sin^2(\pi t / T) \cos(\omega t)$ for $0 \le t \le T$ and $E(t)=0$ otherwise.
$E_0$, $T$ and $\omega$ are the amplitude, length and carrier frequency of the laser pulse, respectively.
Similarly to Ref. \onlinecite{Silva2020}, the radiative emission rate $E_\textrm{R}$ is taken to be proportional to
the expectation value of $\hat{N}$, that is, $E_\textrm{R} \sim \textrm{Tr} (\hat{\rho} \hat{N}) = N(t)$.
In addition, the exciton population is characterized with the expectation value
$\langle \hat{\sigma}^+ \hat{\sigma}^- \rangle(t)$ where $\hat{\sigma}^+$
and $\hat{\sigma}^-$ are raising and lowering operators for the electronic state.

Following earlier work,\cite{Fabri2020234302,Fabri2021a} we again consider the H$_2$CO
molecule and take into account the two singlet electronic states $\textrm{S}_0 ~ (\tilde{\textrm{X}} ~ ^1\textrm{A}_1)$
and $\textrm{S}_1 ~ (\tilde{\textrm{A}} ~ ^1\textrm{A}_2)$.
As already explained in Refs. \onlinecite{Fabri2020234302} and \onlinecite{Fabri2021a}, the 2D($\nu_2$,$\nu_4$) model,
treating the $\nu_2$ (C=O stretch) and $\nu_4$ (out-of-plane)
vibrational modes, provides a physically correct description of H$_2$CO.
Furthermore, the 2D($\nu_2$,$\nu_4$) model, in contrast to one-dimensional quantum-dynamical descriptions,
possesses two internal degrees of freedom required to form LICIs between
polaritonic PESs.\cite{Fabri2020234302,Fabri2021a,Gu20201290,Gu2020a}
We refer to the Supporting Information for further details concerning
the system and technical aspects of the computations.

In what follows, two qualitatively different scenarios under strong light-matter coupling will be presented.
Common to both cases is that the system is initially prepared in a pure state with
$\hat{\rho}(t=0) = | \psi_0 \rangle \langle \psi_0 |$
where $| \psi_0 \rangle$ is the ground state of $\hat{H}_\textrm{cm}$.
Moreover, population transfer occurs primarily to the singly-excited subspace from $| \psi_0 \rangle$ by
the laser pulse and higher-lying polaritonic states have negligible population.
\begin{figure}
\includegraphics[scale=0.5]{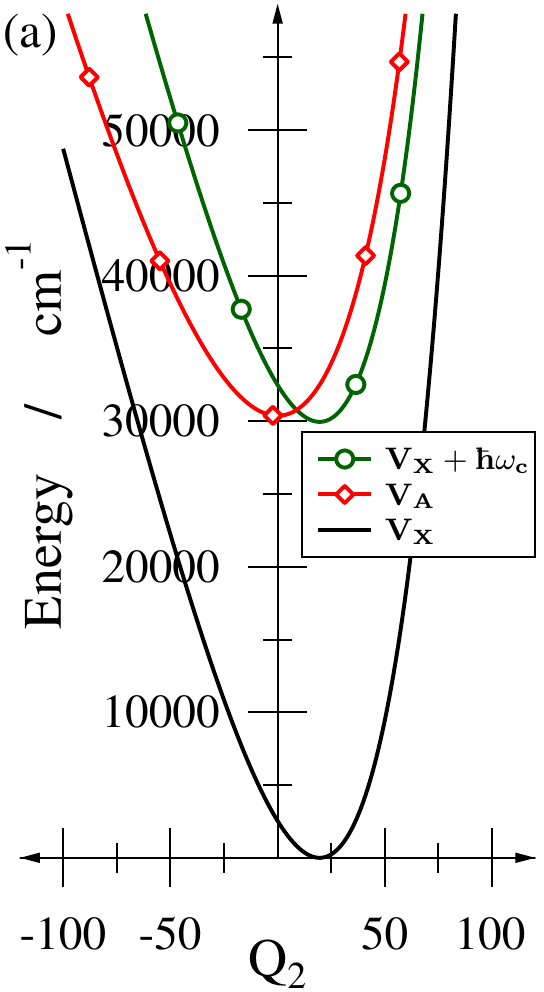}
\includegraphics[scale=0.5]{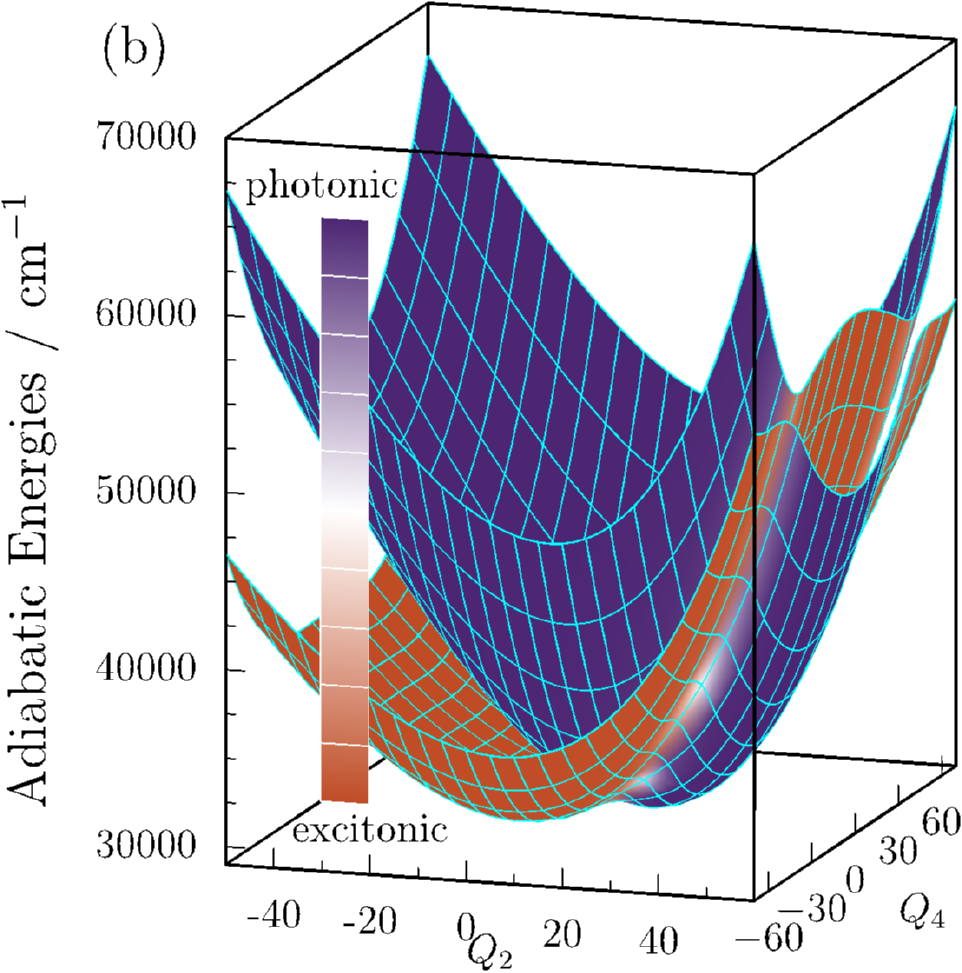}
\includegraphics[scale=0.57]{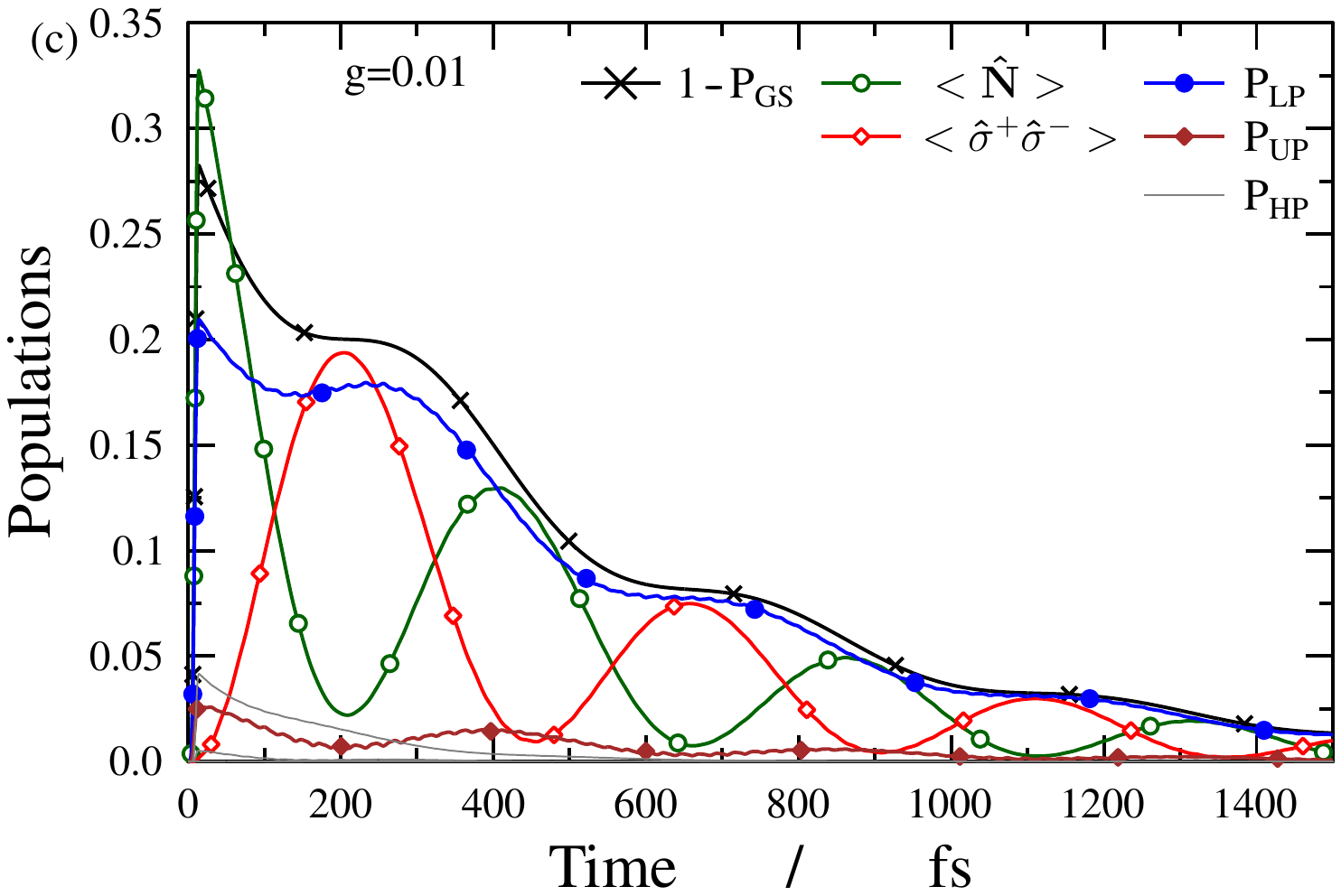}
\includegraphics[scale=0.6]{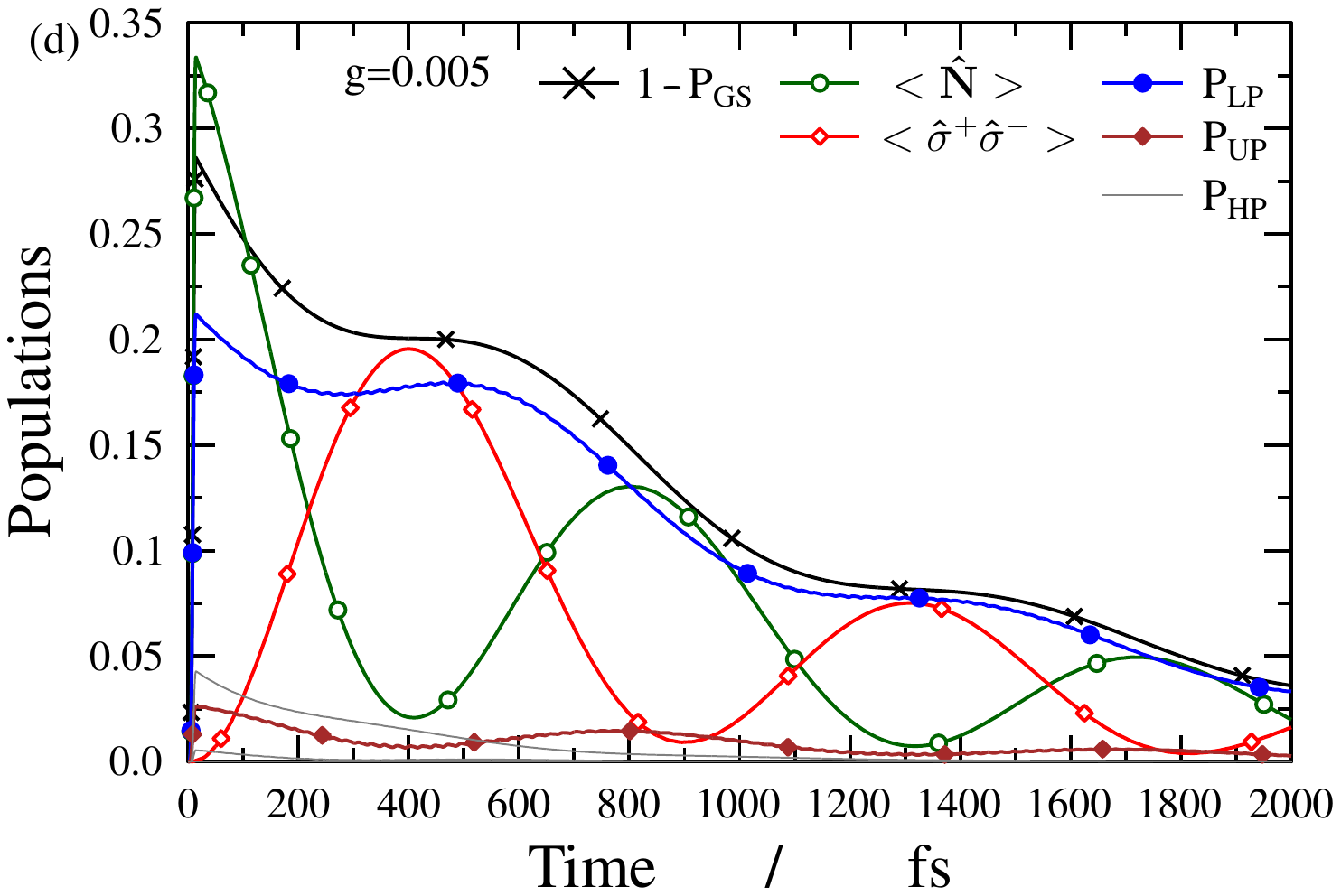}
\caption{\label{fig:case1pop}
		(a) Diabatic potentials ($V_\textrm{X}$, $V_\textrm{A}$ and $V_\textrm{X}+\hbar \omega_\textrm{c}$)
		as a function of the $Q_2$ (C=O stretch) normal coordinate
		(the out-of-plane normal coordinate equals $Q_4 = 0$).
		The cavity wavenumber is $\omega_\textrm{c} = 29957.2 ~ \textrm{cm}^{-1}$.
		(b) Two-dimensional lower (LP) and upper (UP) polaritonic surfaces.
		The cavity wavenumber and coupling strength are 
		$\omega_\textrm{c} = 29957.2 ~ \textrm{cm}^{-1}$ and $g = 0.01 ~ \textrm{au}$, respectively.
		The character of the polaritonic surfaces is indicated by different colors (purple: photonic, orange: excitonic).
		(c-d) Populations of polaritonic states (GS: ground-state (lowest) polariton, HP: higher-lying polaritons)
		and expectation values of the operators $\hat{N}$
		and $\hat{\sigma}^+ \hat{\sigma}^-$ during and after excitation with a $15 ~ \textrm{fs}$ laser pulse
		for $\omega_\textrm{c} = 29957.2 ~ \textrm{cm}^{-1}$, $g = 0.01 ~ \textrm{au}$ (panel c)
		and $g = 0.005 ~ \textrm{au}$ (panel d).
		The emission is proportional to the expectation value of $\hat{N}$, $N(t)$.}
\end{figure}
\begin{figure}
\includegraphics[scale=0.65]{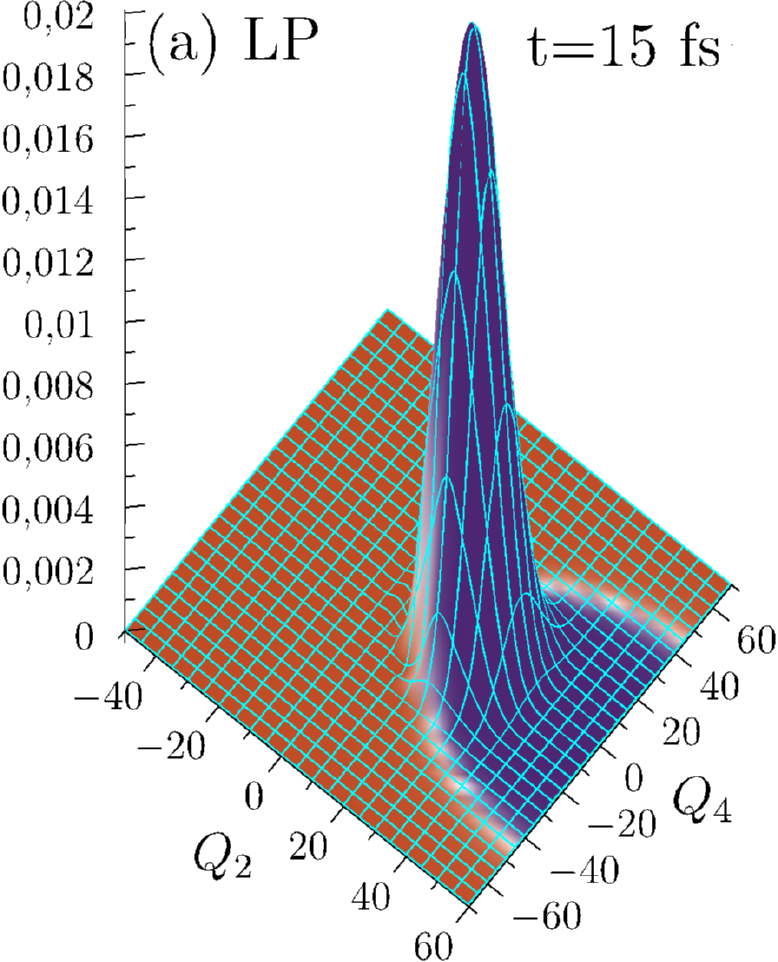}
\includegraphics[scale=0.65]{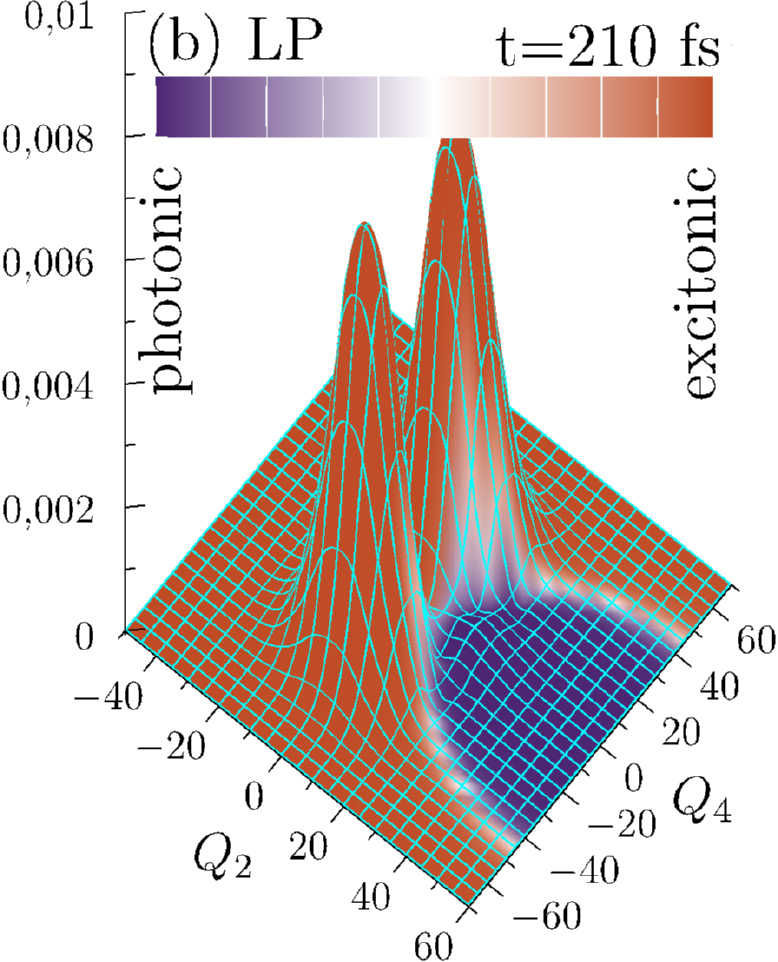}
\includegraphics[scale=0.65]{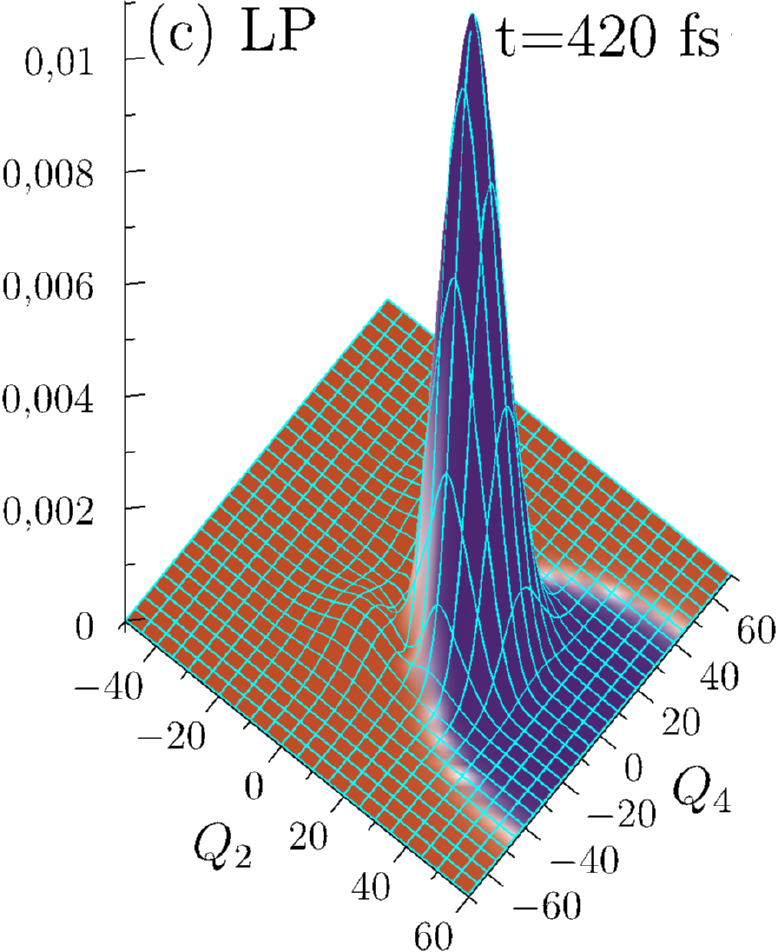}
\caption{\label{fig:case1wp}
		Probability density figures for the lower polaritonic (LP) surface (time is indicated in the figure captions)
		with $\omega_\textrm{c} = 29957.2 ~ \textrm{cm}^{-1}$ and $g = 0.01 ~ \textrm{au}$.
		The laser pulse selectively populates the lower polariton and oscillations
		in the $N(t)$ curve (see panel c in Figure \ref{fig:case1pop} for $g = 0.01 ~ \textrm{au}$)
		can be attributed to the motion of the wavepacket between the photonic
		(purple) and excitonic (orange) regions of the LP surface.}
\end{figure}

In the first case the cavity wavenumber and coupling strength equal
$\omega_\textrm{c} = 29957.2 ~ \textrm{cm}^{-1}$ and 
$g = 0.01 ~ \textrm{au}$ ($\gamma_\textrm{c} = 10^{-4} ~ \textrm{au}$) or 
$g = 0.005 ~ \textrm{au}$ ($\gamma_\textrm{c} = 5 \cdot 10^{-5} ~ \textrm{au}$), respectively, while
the LICI is located at $Q_2 = 8.84$ and $Q_4 = 0$ at an energy corresponding to $30776.1 ~ \textrm{cm}^{-1}$.
The relevant diabatic and polaritonic PESs (for $g = 0.01 ~ \textrm{au}$)
are depicted in Figure \ref{fig:case1pop} (panels a and b).
If the cavity mode is pumped with a laser pulse of $\omega = 30000 ~ \textrm{cm}^{-1}$,
$T = 15 ~ \textrm{fs}$ and $E_0 = 3.77 \cdot 10^{-3} ~ \textrm{au}$
(equivalent to an intensity of $I = 5 \cdot 10^{11} ~ \textrm{W} / \textrm{cm}^{2}$),
the ensuing quantum dynamics is basically adiabatic and takes place almost 
entirely on the LP PES which has excitonic character on the left
and photonic on the right as indicated in Figure \ref{fig:case1pop} (panel b).
After $t = 15 ~ \textrm{fs}$, about $20 \%$ of the initial ground-state
population is transferred primarily to the photonic part of the LP PES and the
nuclear wavepacket starts to move back and forth between the photonic and excitonic regions of LP adiabatically,
which is clearly reflected by oscillations in the emission (Figure \ref{fig:case1pop}, panels c and d).
The lack of nonadiabatic transition between LP and UP is also apparent in the time-dependent populations in
Figure \ref{fig:case1pop} (panels c and d).
For $g = 0.01 ~ \textrm{au}$ and $t = 210 ~ \textrm{fs}$,
the wavepacket is mainly localized in the excitonic part of the LP PES,
and as a consequence, the emission and $\langle \sigma^+ \sigma^- \rangle(t)$
show a local minimum and maximum, respectively.
Other local minima and maxima of the emission and $\langle \sigma^+ \sigma^- \rangle(t)$
can be rationalized along similar lines.
The motion of the wavepacket between the photonic and excitonic parts of the LP PES is clearly visible
in the probability density figures of Figure \ref{fig:case1wp}.
The oscillatory behavior of the emission and $\langle \sigma^+ \sigma^- \rangle(t)$ continues over time
with a decreasing amplitude due to radiative relaxation.
This trend is also reflected in the temporal behavior of the nuclear probability density.
\begin{figure}
\includegraphics[scale=0.45]{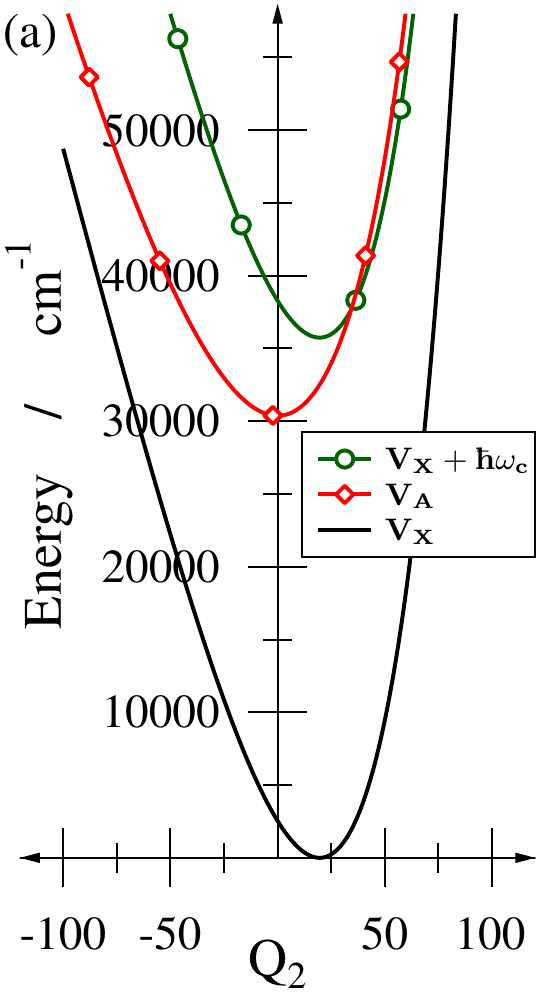}
\includegraphics[scale=0.5]{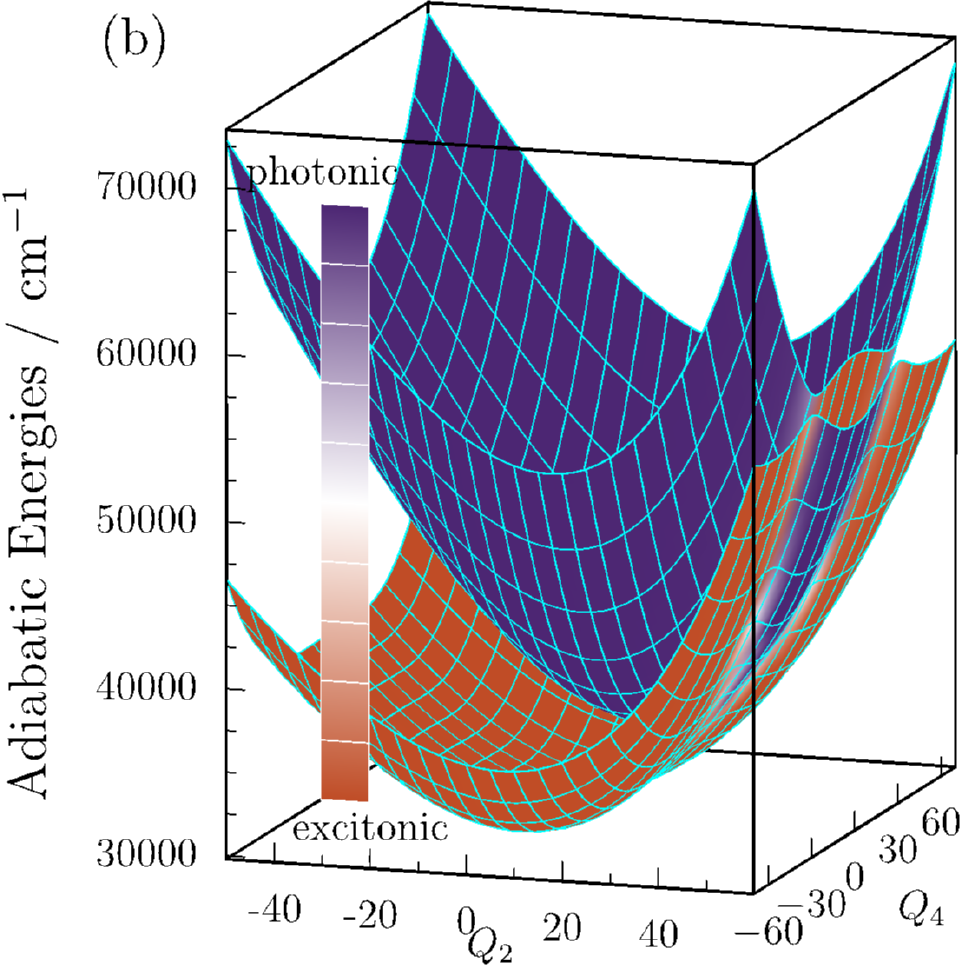}
\includegraphics[scale=0.525]{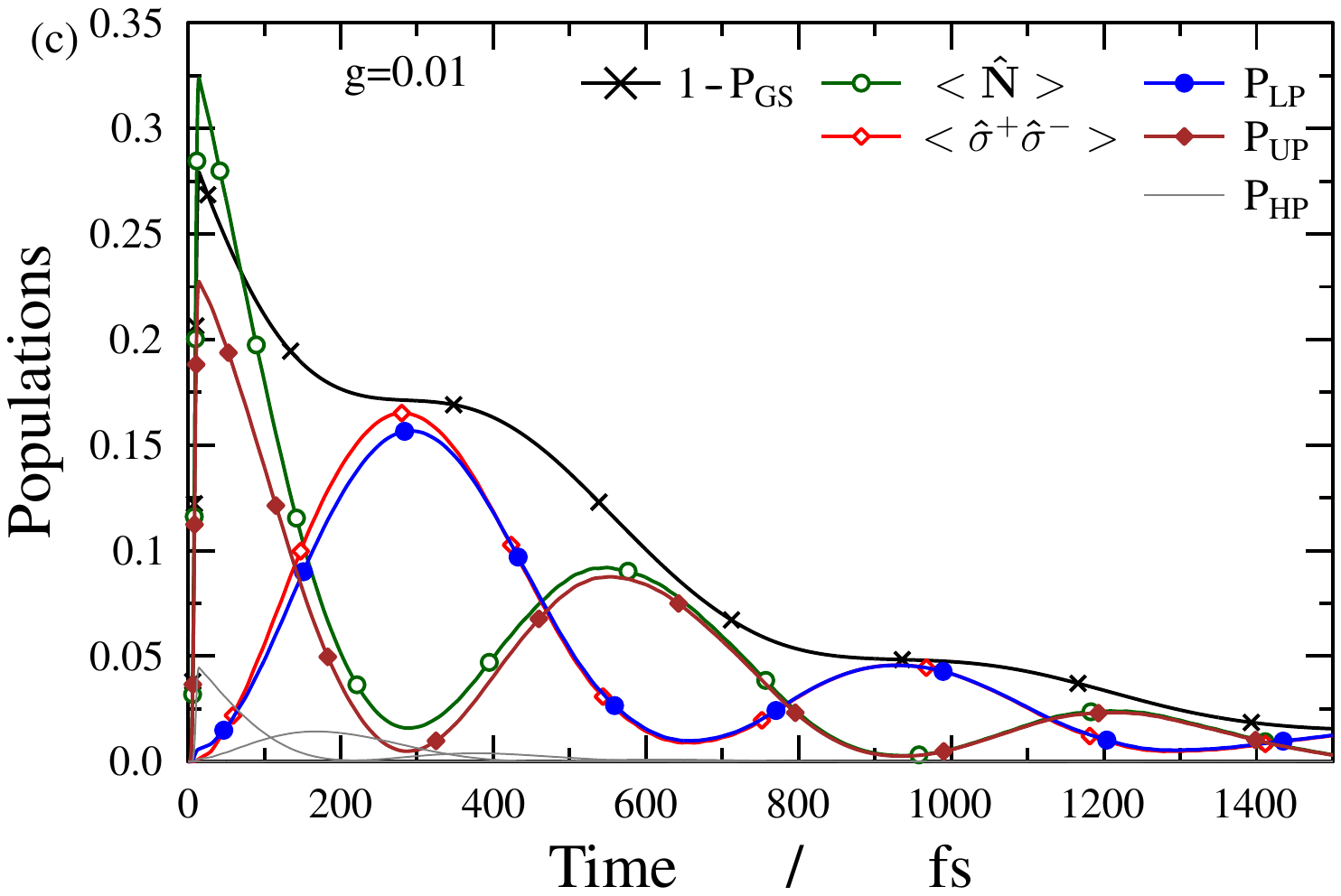}
\includegraphics[scale=0.525]{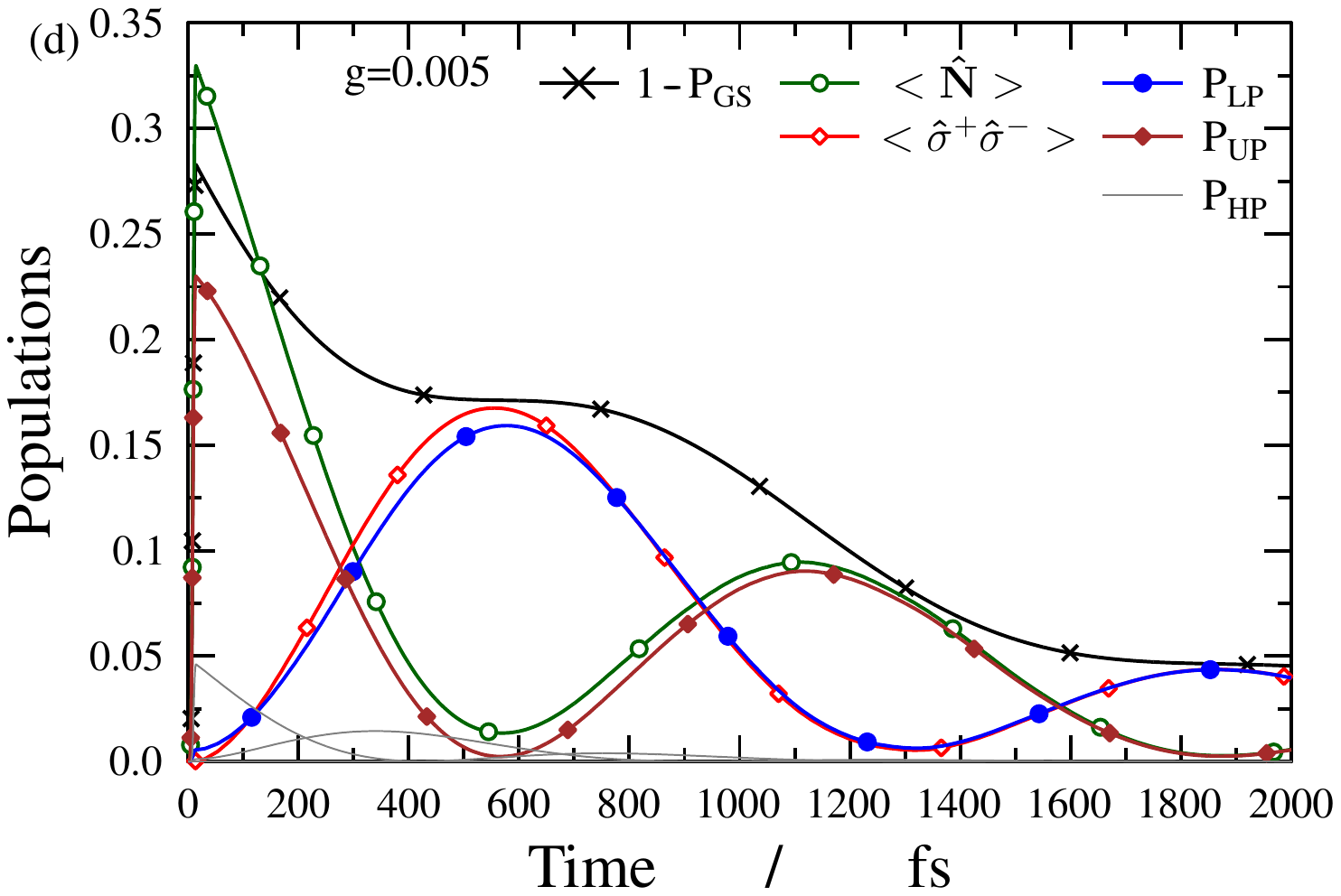}
\includegraphics[scale=0.525]{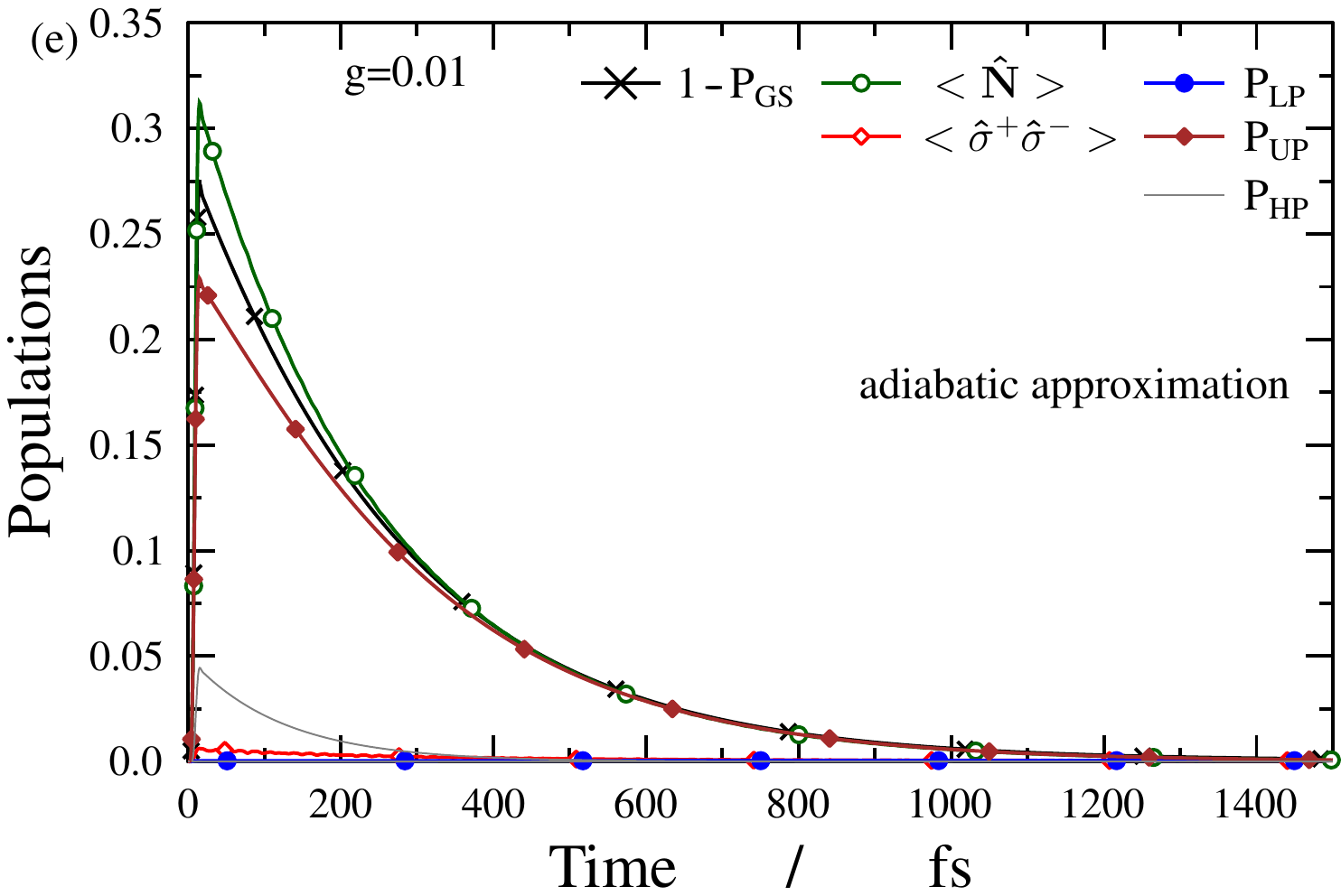}
\caption{\label{fig:case2pop}
		(a) Diabatic potentials ($V_\textrm{X}$, $V_\textrm{A}$ and $V_\textrm{X}+\hbar \omega_\textrm{c}$)
		as a function of the $Q_2$ (C=O stretch) normal coordinate (the out-of-plane normal coordinate equals $Q_4 = 0$).
		The cavity wavenumber is $\omega_\textrm{c} = 35744.8 ~ \textrm{cm}^{-1}$.
		(b) Two-dimensional lower (LP) and upper (UP) polaritonic surfaces. The cavity wavenumber and coupling strength are 
		$\omega_\textrm{c} = 35744.8 ~ \textrm{cm}^{-1}$ and $g = 0.01 ~ \textrm{au}$, respectively.
		The character of the polaritonic surfaces is indicated by different colors (purple: photonic, orange: excitonic).
		(c-d) Populations of polaritonic states (GS: ground-state (lowest) polariton, HP: higher-lying polaritons)
		and expectation values of the operators $\hat{N}$ and $\hat{\sigma}^+ \hat{\sigma}^-$ during and after excitation
		with a $15 ~ \textrm{fs}$ laser pulse for $\omega_\textrm{c} = 35744.8 ~ \textrm{cm}^{-1}$, $g = 0.01 ~ \textrm{au}$ (panel c)
		and $g = 0.005 ~ \textrm{au}$ (panel d). The emission is proportional to the expectation value of $\hat{N}$, $N(t)$.
		(e) Same as (c) ($\omega_\textrm{c} = 35744.8 ~ \textrm{cm}^{-1}$, $g = 0.01 ~ \textrm{au}$) using the adiabatic approximation.
		Note the stark contrast between the adiabatic and exact results:
		the adiabatic $N(t)$ curve follows a simple exponential decay while
		the exact $N(t)$ curve shows an oscillatory behavior.}
\end{figure}
\begin{figure}
\includegraphics[scale=0.65]{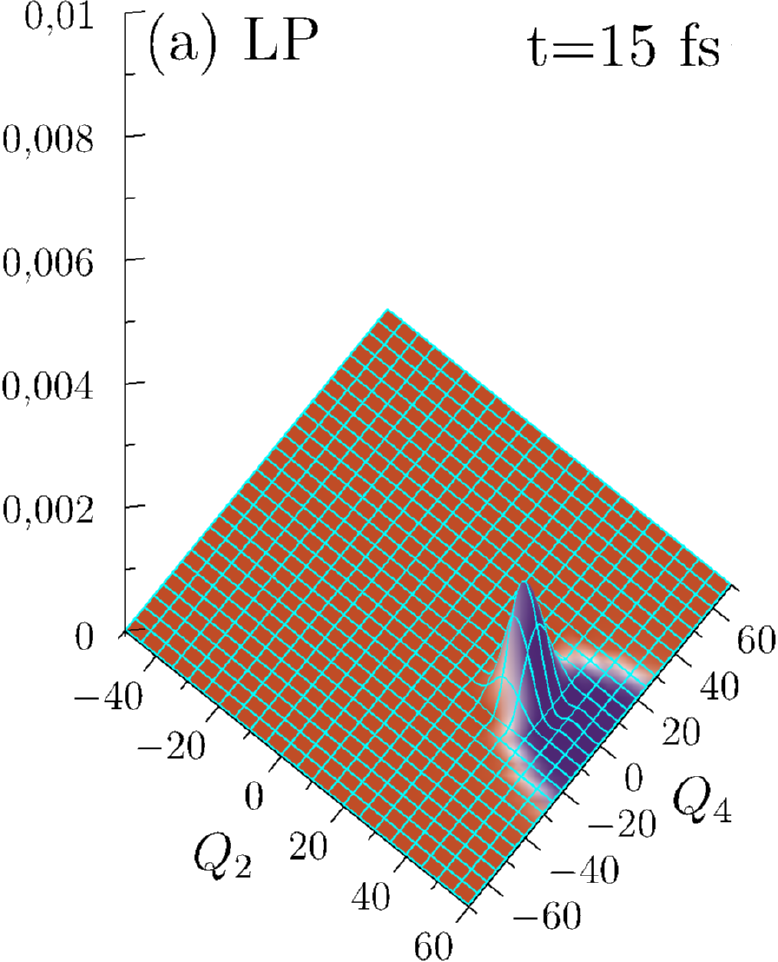}
\includegraphics[scale=0.65]{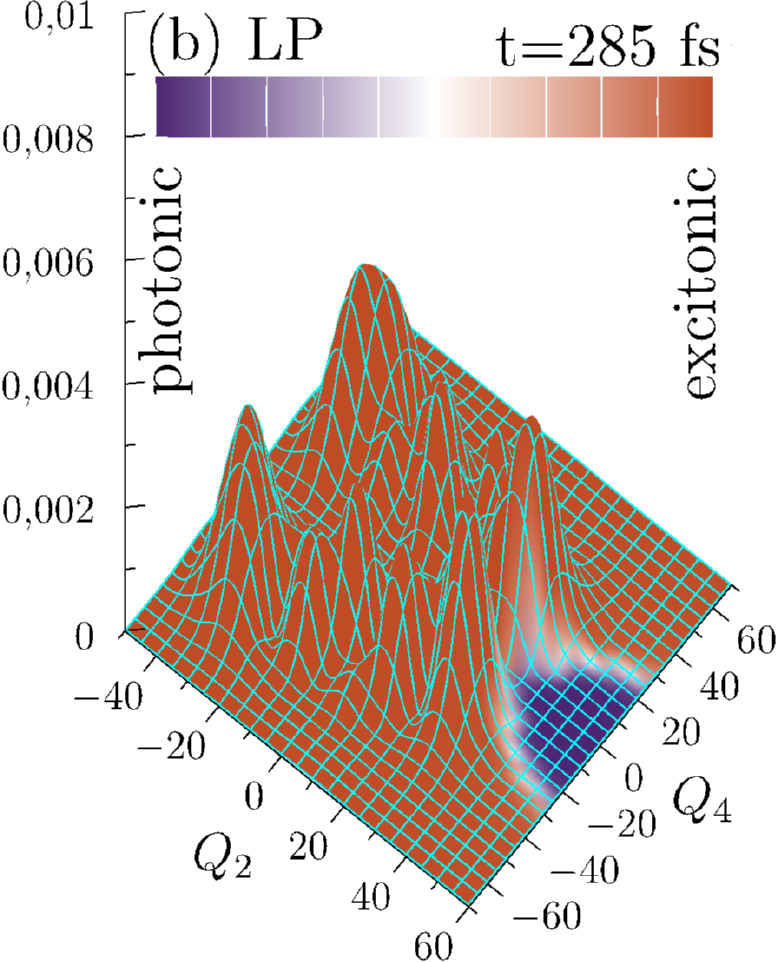}
\includegraphics[scale=0.65]{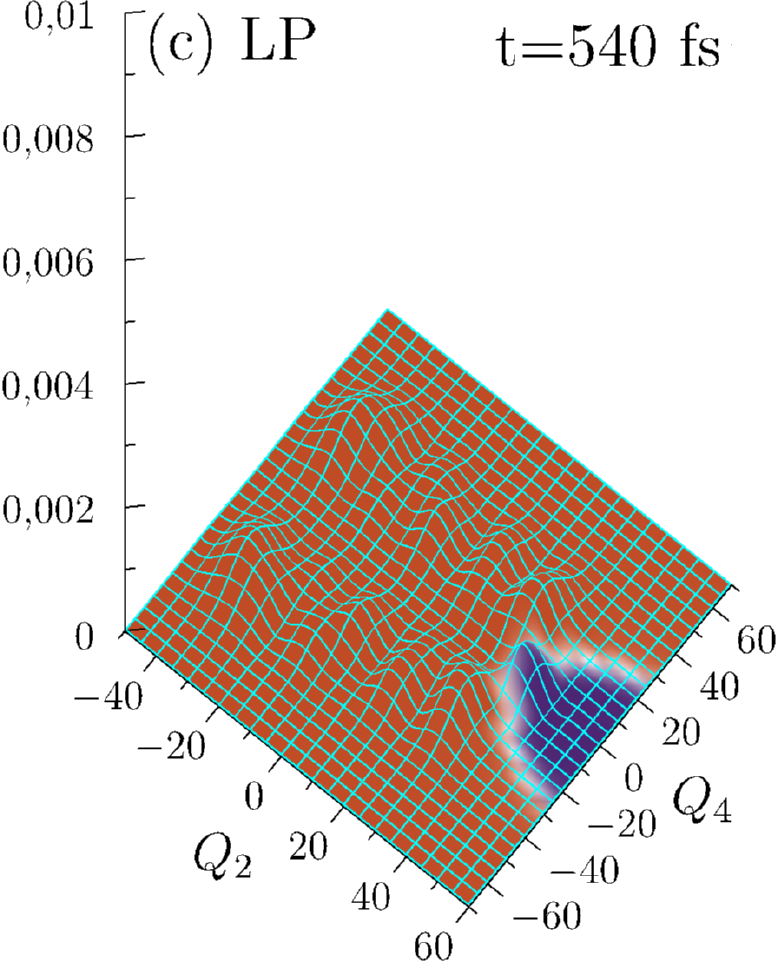}
\includegraphics[scale=0.65]{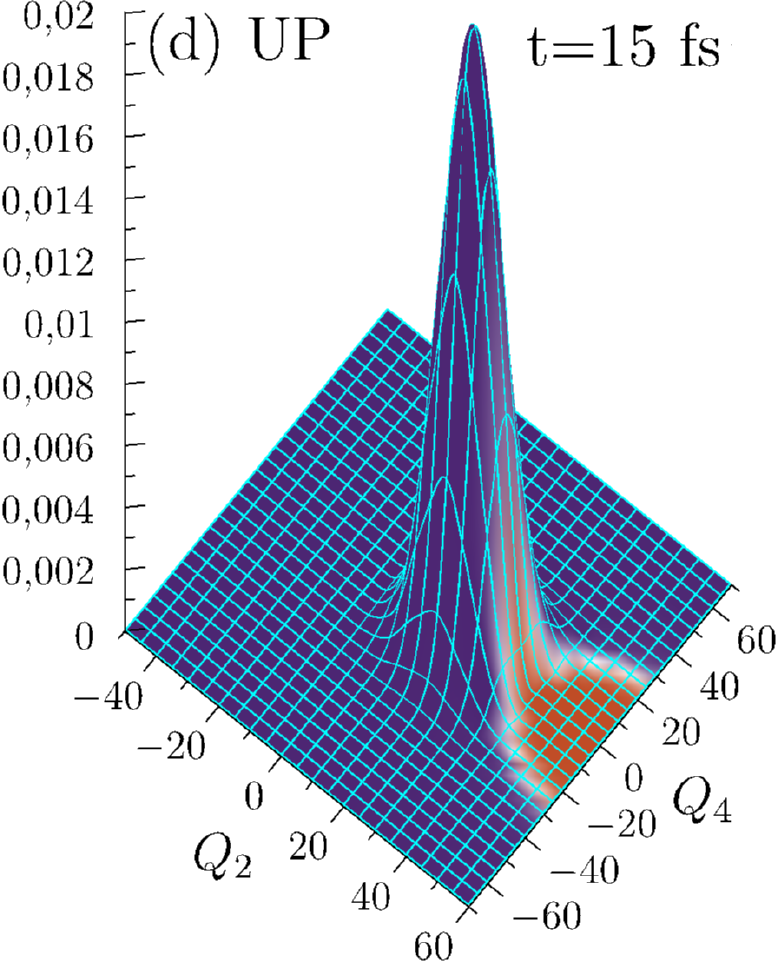}
\includegraphics[scale=0.65]{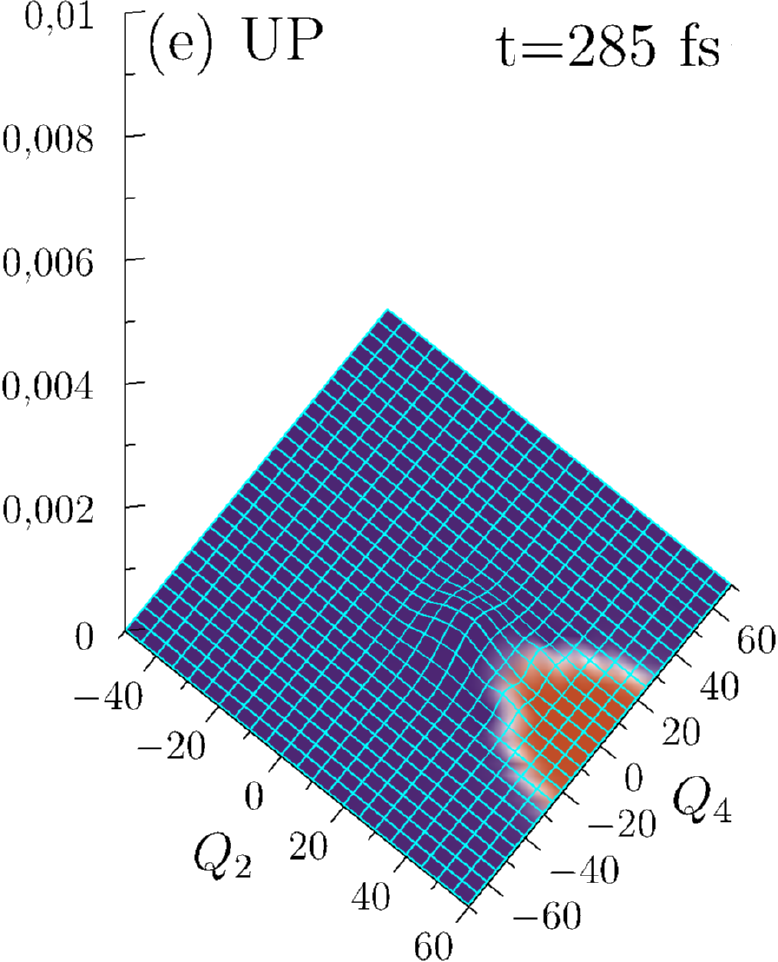}
\includegraphics[scale=0.65]{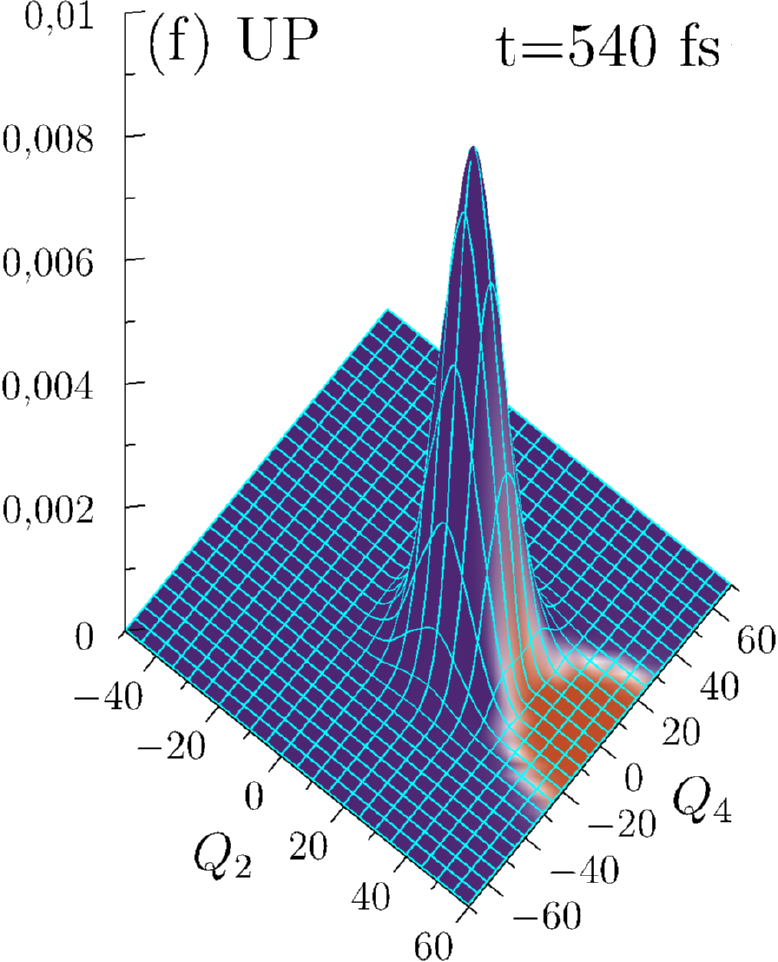}
\caption{\label{fig:case2wp}
		Probability density figures for the lower (LP, panels a-c) and upper (UP, panels d-f) polaritonic surfaces
		with $\omega_\textrm{c} = 35744.8 ~ \textrm{cm}^{-1}$ and $g = 0.01 ~ \textrm{au}$ (time is indicated in the figure captions).
		Oscillations in the exact $N(t)$ curve (see panel c in Figure \ref{fig:case2pop} for $g = 0.01 ~ \textrm{au}$) are due to
		nonadiabatic population transfer between the photonic (purple) region of the UP and excitonic (orange) region of the LP surfaces.}
\end{figure}

The situation is entirely different if the cavity parameters are set to $\omega_\textrm{c} = 35744.8 ~ \textrm{cm}^{-1}$
and
$g = 0.01 ~ \textrm{au}$ ($\gamma_\textrm{c} = 10^{-4} ~ \textrm{au}$) or 
$g = 0.005 ~ \textrm{au}$ ($\gamma_\textrm{c} = 5 \cdot 10^{-5} ~ \textrm{au}$).
In this case the LICI is shifted to $Q_2 = 34.28$ and $Q_4 = 0$,
while the energy of the LICI corresponds to $37650.8 ~ \textrm{cm}^{-1}$.
The relevant diabatic and polaritonic PESs (for $g = 0.01 ~ \textrm{au}$)
are depicted in Figure \ref{fig:case2pop} (panels a and b).
As can be seen in Figure \ref{fig:case2pop}, the shape of the UP PES around its minimum is approximately
identical to that of $V_\textrm{X}$. The same argument applies to the lowest (GS) polaritonic PES.
Therefore, the lowest adiabatic eigenstates of both GS and UP are images of the X vibrational ground state.
With this in mind, we have employed a laser pulse ($\omega = 36000 ~ \textrm{cm}^{-1}$,
$T = 15 ~ \textrm{fs}$ and $E_0 = 3.77 \cdot 10^{-3} ~ \textrm{au}$) tailored  to generate a copy
of the vibrational ground state on the UP PES.
This is apparent in the probability density for $t=15 ~ \textrm{fs}$
(see panel d of Figure \ref{fig:case2wp} for $g = 0.01 ~ \textrm{au}$)
which also shows that the wavepacket is almost fully localized in the photonic region of the UP PES.
As the UP wavepacket has nonzero amplitude around the LICI, nonadiabatic population transfer
from UP to LP sets in, clearly visible in the polaritonic populations in Figure \ref{fig:case2pop} (panels c and d).
At the end of the laser pulse ($t = 15 ~ \textrm{fs}$) the LP PES possesses a negligible
amount of population for both $g$ values considered.
For $g = 0.01 ~ \textrm{au}$, the UP population reaches a local minimum around $t = 285 ~ \textrm{fs}$
and so does the emission, while the LP population shows a local maximum. Thereafter, both the
UP population and emission increase again, reaching a local maximum around $t = 540 ~ \textrm{fs}$.
One can also notice in the probability density figures of Figure \ref{fig:case2wp}
that the LP (UP) wavepacket remains mainly in the excitonic (photonic) region of the LP (UP) PESs.
These observations imply that oscillations of the emission signal can be mainly attributed to
nonadiabatic population transfer between the photonic UP and excitonic LP PES regions.
The emission and UP population curves follow each other and the oscillation period is dictated by the LICI.
For later times the efficiency of the nonadiabatic population transfer decreases,
which is not surprising, since conical intersections are known to modify the short-time dynamics.\cite{Domcke2004}

We have also performed computations neglecting nonadiabatic coupling between polaritonic PESs
(adiabatic approximation) with $\omega_\textrm{c} = 35744.8 ~ \textrm{cm}^{-1}$,
$g = 0.01 ~ \textrm{au}$ ($\gamma_\textrm{c} = 10^{-4} ~ \textrm{au}$) and
$g = 0.005 ~ \textrm{au}$ ($\gamma_\textrm{c} = 5 \cdot 10^{-5} ~ \textrm{au}$) (second case).
The polaritonic populations and expectation values for $g = 0.01 ~ \textrm{au}$ are depicted in
Figure \ref{fig:case2pop} (panel e).
It is striking that both the adiabatic UP population and emission show a simple exponential decay after $t = 15 ~ \textrm{fs}$.
This is to be fully contrasted with previous exact results which account for nonadiabatic effects.
Note that similar conclusions can be drawn for the $g = 0.005 ~ \textrm{au}$ case (see the Supporting Information
for populations and expectation values for $g = 0.005 ~ \textrm{au}$ in the adiabatic approximation).
Moreover, nonlinear fit of $f(t) = a \exp(-\gamma t)$ to the adiabatic emission data points with $t \ge 15 ~ \textrm{fs}$
results in $\gamma = 9.83 \cdot 10^{-5} ~ \textrm{au}$ ($g = 0.01 ~ \textrm{au}$) and 
$\gamma = 4.91 \cdot 10^{-5} ~ \textrm{au}$ ($g = 0.005 ~ \textrm{au}$)
which agree well with $\gamma_\textrm{c} = 10^{-4} ~ \textrm{au}$ ($g = 0.01 ~ \textrm{au}$) and
$\gamma_\textrm{c} = 5 \cdot 10^{-5} ~ \textrm{au}$ ($g = 0.005 ~ \textrm{au}$), respectively.
As explained in the previous section, the laser pulse selectively populates the lowest UP
adiabatic eigenstate which is localized in the photonic region of the UP PES. Since the adiabatic approximation 
neglects nonadiabatic coupling between the polaritonic PESs, no population transfer is possible from UP to LP
and we get an exponentially decaying emission signal (see the Supporting Information for an analytical derivation).
Population and expectation value figures on logarithmic scale
are provided in the Supporting Information, also supporting our conclusions.

Having discussed both the exact and adiabatic results for the second case, one can conclude that
the oscillatory behavior of the emission can be attributed to nonadiabatic population
transfer between the UP and LP PESs in this case.
This finding serves as a dynamical fingerprint of the LICI, supplementing spectroscopic\cite{Fabri2021a}
and topological\cite{21BaUmFa} LICI-related effects observed earlier in H$_2$CO.
Since the laser pulse dominantly populates the lowest UP eigenstate,
the probability density is stationary and the wavepacket will not move between the photonic and
excitonic regions of the UP PES. Based on this argument, one would not expect any oscillations in the emission.
However, as the UP wavepacket has nonzero amplitude around the LICI, nonadiabatic population
exchange occurs between the UP and LP PESs, which manifests in an oscillatory emission signal.
We stress that the second case can be clearly distinguished from the
first case where the emission also displays oscillatory behavior.
In the first case, similarly to Ref. \onlinecite{Silva2020}, oscillations
are obviously the consequence of wavepacket motion between the photonic and excitonic regions
of a single polaritonic PES.
We emphasize that different mechanisms of the first and second cases are due to the initial setup.
Namely, the cavity frequency (determining the LICI position)  and the laser frequency
(dictating the extent to which polaritonic PESs are populated by the laser pulse) are different.
In view of this, the origin of oscillations in the emission can be unequivocally identified in both cases presented.

The $g = 0.01 ~ \textrm{au}$ and $g = 0.005 ~ \textrm{au}$ coupling strength values used throughout this work
correspond to the $1-10 ~ \textrm{nm}^3$ mode volume range. Further results with higher ($g = 0.1 ~ \textrm{au}$)
and lower ($g = 0.002 ~ \textrm{au}$) couplings are provided in the Supporting Information.
Based on the analysis of computations with different $g$ values we can conclude that essentially the same effects
(wavepacket motion on a single polaritonic  PES or nonadiabatic population transfer between polaritonic PESs)
can be seen for different coupling strengths, which demonstrates the robustness of the suggested method.
It is also conspicuous in the emission signals that reducing the coupling strength implies slower dynamics
and longer oscillation periods in the emission. A direct consequence of this observation is that for lower coupling
strengths lower $\gamma_\textrm{c} $ values (in other words, cavities with higher quality factors) are needed if
one wants to see any oscillations in the emission signal before relaxation to the ground state occurs.
In return, local minima and maxima of the emission become more pronounced
for lower values of $g$, which is readily explained by sharper boundaries between the excitonic and photonic
regions of polaritonic PESs for weaker couplings.

To conclude, we have suggested a method to probe light-induced conical
intersections in a realistically-described polyatomic molecule by following the
time-dependent radiative emission in a lossy plasmonic nano-cavity.
The emission substitutes the probe pulse used by standard pump-probe experiments.
By monitoring nonadiabatic wavepacket dynamics we could clearly identify
ultrafast population transfer through the LICI, manifesting in the
oscillatory behavior of the emission. Although the initial setup in our specific case 
has enabled to clearly identify the fingerprint of nonadiabatic population transfer,
generally the emission is shaped by the combined effect of wavepacket
motion on individual polaritonic PESs and population flow between polaritonic PESs,
which requires further investigation.
We hope that our results will stimulate combined laser-cavity
experiments and the extension of theory to treat more
complex molecules in a cavity. The proposed procedure can be further generalized to
the situation where a natural CI is also present in the system and
nonadiabatic dynamics is governed by the interplay between natural and
light-induced nonadiabatic phenomena. The latter can, however, be
controlled and used to modify the dynamics, for instance, to suppress or
enhance the impact of the former.

\clearpage

\begin{acknowledgments}
The authors are indebted to NKFIH for financial support (Grant K128396).
Lorenz S. Cederbaum is gratefully acknowledged for insightful discussions.
\end{acknowledgments}

\bibliography{polaritonic_clock_arxiv}

\newpage

\begin{center}
{\Large Supporting Information}
\end{center}

\section{The H$_{2}$CO molecule and technical details of the computations}
\begin{figure}[hbt!]
   \begin{center}
      \includegraphics[scale=0.3]{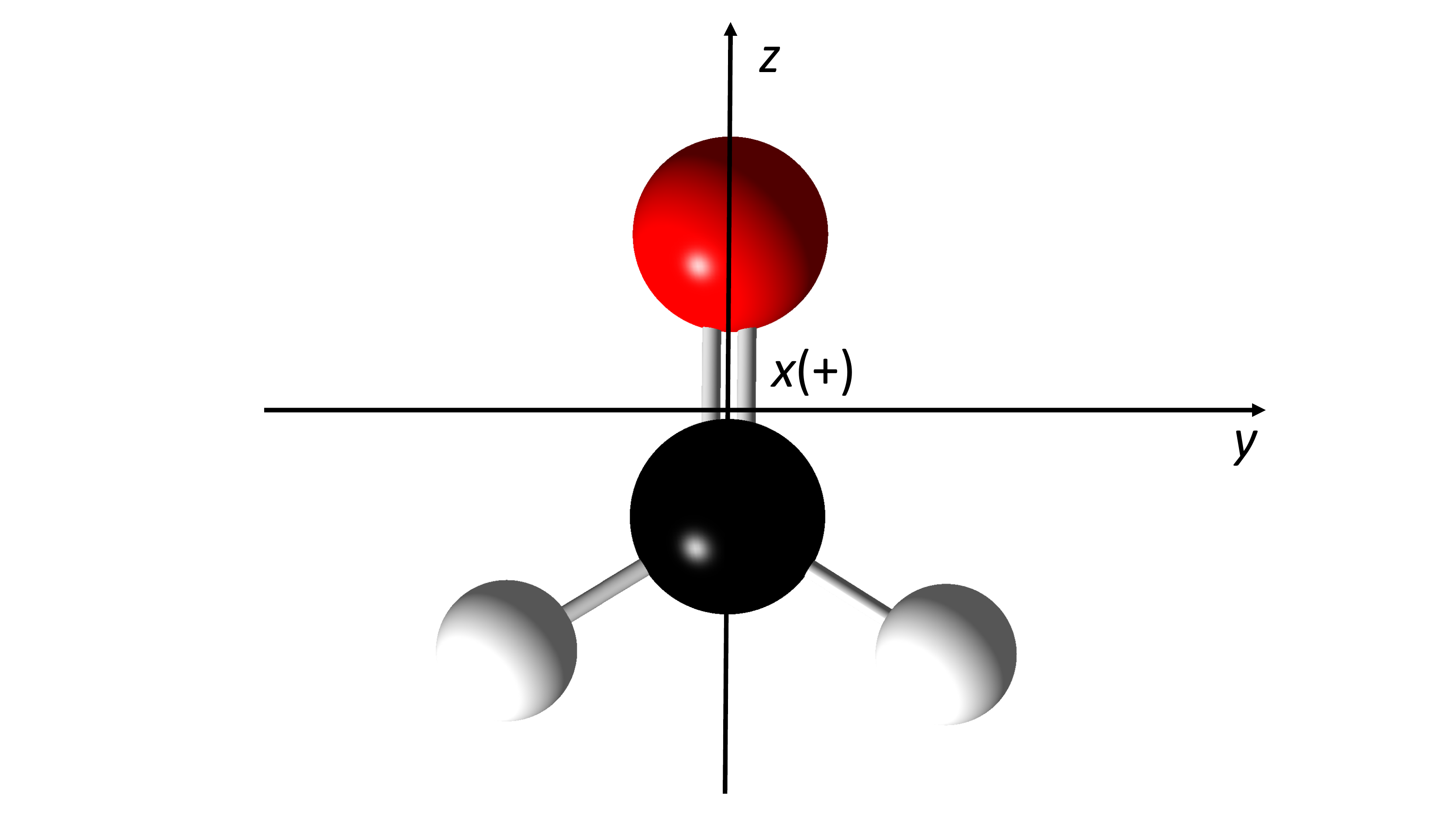}
      \includegraphics[scale=0.5]{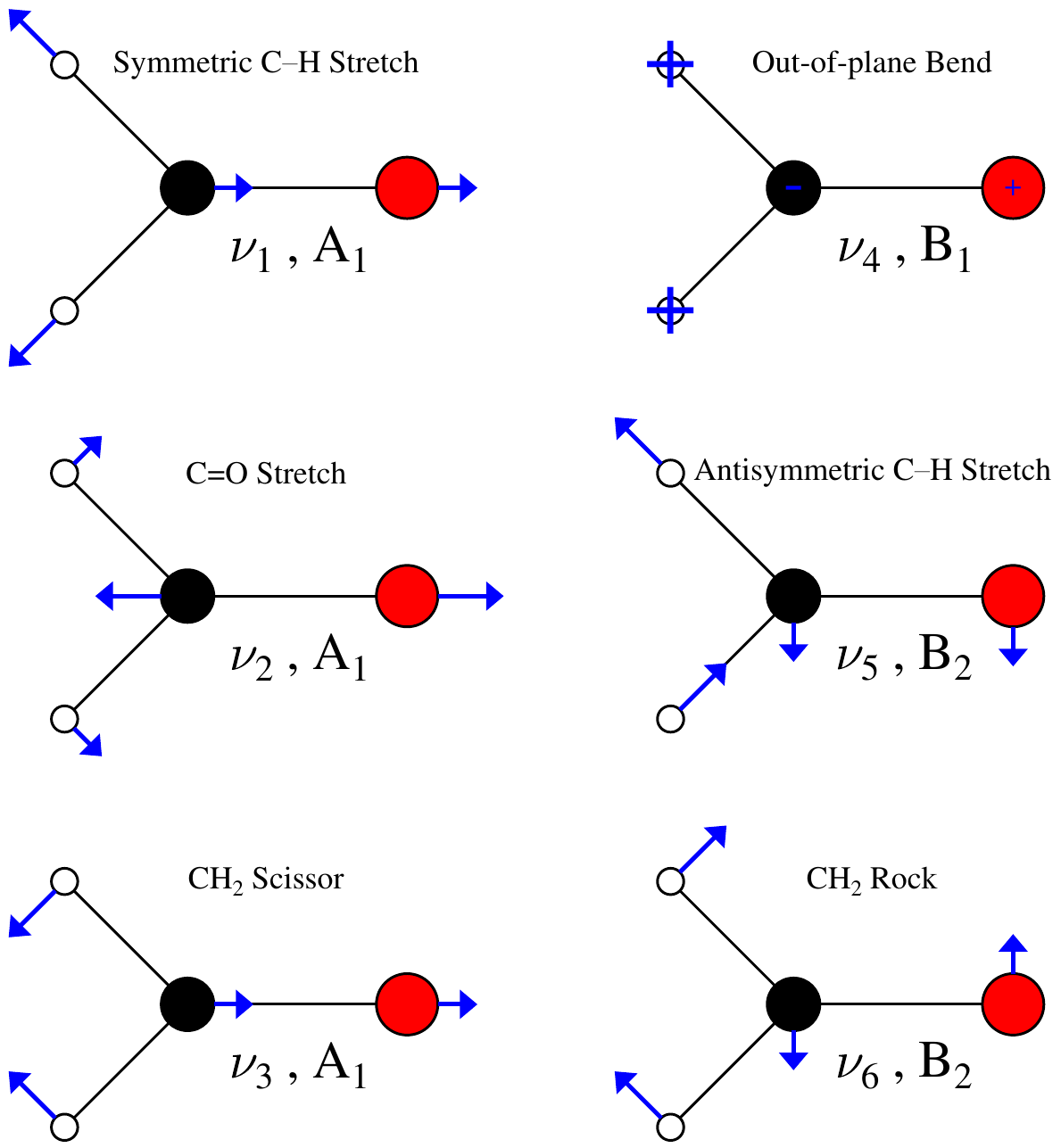}
   \end{center}
   \caption{\label{fig:structure_modes}
        Equilibrium structure of the H$_{2}$CO molecule (left panel) in its ground electronic state (X) with
   	definition of the body-fixed Cartesian axes (the $x$ axis is directed outwards, as indicated by the $+$ sign)
	and vibrational normal modes of the H$_{2}$CO molecule (right panel).}
\end{figure}

H$_{2}$CO has a planar equilibrium structure ($C_{2v}$ point-group symmetry) in its ground electronic state
($\tilde{\textrm{X}} ~ ^1\textrm{A}_1$), shown in Figure \ref{fig:structure_modes} (left panel).
In the excited electronic state ($\tilde{\textrm{A}} ~ ^1\textrm{A}_2$) H$_{2}$CO has two equivalent nonplanar equilibrium structures
($C_{s}$ point-group symmetry) which are connected by a planar transition state structure ($C_{2v}$ point-group symmetry).
The six vibrational normal modes of H$_{2}$CO are depicted in Figure \ref{fig:structure_modes} (right panel).
Due to symmetry, the transition dipole moment (TDM) vector vanishes at any geometry of $C_{2v}$ symmetry.
Moreover, H$_{2}$CO does not have any first-order nonadiabatic coupling between the X and A electronic states
around its equilibrium geometry, which implies that light-induced nonadiabatic effects can be unambiguously separated
from natural nonadiabatic effects.

The 2D($\nu_2$,$\nu_4$) vibrational model of H$_2$CO was defined by evaluating 
the normal coordinates corresponding to the planar transition state structure of the excited electronic state
and setting values of the inactive normal coordinates ($Q_1$, $Q_3$, $Q_5$, $Q_6$) to zero.
The 2D($\nu_2$,$\nu_4$) potential energy surfaces (PESs) ($V_\textrm{X}$ and $V_\textrm{A}$)
and the TDM surface were calculated as a function of the
$Q_2$ (C=O stretch) and $Q_4$ (out-of-plane) normal coordinates at the CAM-B3LYP/6-31G* level of theory.
2D PES and TDM functions were generated by interpolating the ab initio PES and TDM points.
In the 2D($\nu_2$,$\nu_4$) model the TDM vector is always perpendicular to the permanent dipole moment
vector of both electronic states and only the body-fixed $y$ component of the TDM vector can be nonzero.
In all computations the polarization vector of the cavity field coincides with the body-fixed $y$ axis.
As the TDM vanishes for $Q_4=0$ due to symmetry, the lower and upper polaritonic PESs cross each other
(in other words, a light-induced conical intersection is formed)
if $V_\textrm{X} + \hbar \omega_\textrm{c} = V_\textrm{A}$ (crossing of diabatic PESs, $\omega_\textrm{c}$
is the cavity frequency) and $Q_4=0$ (zero coupling).

The Lindblad master equation was solved numerically using the combination of 2D direct-product discrete variable representation
basis functions and Fock states of the cavity mode $| n \rangle$ with $n=0,1,2$.
The polaritonic PESs were obtained by diagonalizing the potential energy part of the Hamiltonian at each nuclear geometry
(see Eq. (2) of the manuscript).
The approximate adiabatic computations were carried out by transforming the Lindblad master equation to the adiabatic
representation and neglecting nonadiabatic coupling between polaritonic PESs.
This requires the computation of a unitary transformation matrix $\bf{U}$ which diagonalizes the potential energy matrix
and transforms the Lindblad master equation to the adiabatic representation. $\bf{U}$ contains the eigenvectors
of the potential energy matrix in its columns.

\section{Time-dependence of the expectation value of the photon number operator in the adiabatic approximation}

In what follows, a simple model is presented to justify the exponential decay  of the expectation value of the photon number operator
in the adiabatic approximation for $\omega_\textrm{c} = 35744.8 ~ \textrm{cm}^{-1}$ (second case in the manuscript).
With this cavity wavenumber the shapes of the upper polaritonic (UP) and ground-state polaritonic (GS) PESs are
essentially identical to that of the ground-state PES ($V_\textrm{X}$) around their minima. 
As a consequence, the lowest adiabatic eigenstates of the UP ($| \varphi_1 \rangle$) 
and GS ($| \varphi_0 \rangle$) PESs can be well approximated as
\begin{align}
	| \varphi_0 \rangle &= | \textrm{X} \rangle | 0 \rangle | \varphi \rangle \\
	| \varphi_1 \rangle &= | \textrm{X} \rangle | 1 \rangle | \varphi \rangle \nonumber
\end{align}
where $| \varphi \rangle$ is the vibrational ground state of the X electronic state.
$E_0$ and $E_1$ denote the respective energy levels with $E_1-E_0 = \hbar \omega_\textrm{c}$.
If the system interacts with a laser pulse whose carrier frequency is in (near) resonance with $\omega_\textrm{c}$,
it is sufficient to consider the states $| \varphi_0 \rangle$ and $| \varphi_1 \rangle$ as basis states
under the conditions of the second case of the manuscript.
In other words, $| \varphi_1 \rangle$ is virtually the only bright state around $E_1$
coupled to $| \varphi_0 \rangle$ by the operator
\begin{equation}
	\hat{V}(t) = -\mu_\textrm{c} E(t) (\hat{a}^\dagger + \hat{a})
\end{equation}
which describes the interaction of the system with a laser pulse of length $T$.
The validity of the two-state approximation in this particular case has been confirmed by numerical computations.
$\hat{V}(t)$ gives rise to the coupling matrix element
\begin{equation}
	\langle \varphi_0 | \hat{V}(t) | \varphi_1 \rangle = -\mu_\textrm{c} E(t).
\end{equation}
It is important to note that $\hat{V}(t)$ has approximately the same form in both the diabatic and adiabatic representations
in the region where $| \varphi_0 \rangle$ and $| \varphi_1 \rangle$ have nonzero amplitude.
The system is initially ($t=-T$) in the ground state, characterized by the density matrix
\begin{equation}
	\rho(-T) = 
	\begin{bmatrix}
            1 & 0 \\
            0 & 0
        \end{bmatrix}
\end{equation}
which evolves into
\begin{equation}
	\rho(0) = 
	\begin{bmatrix}
            \rho_{00}(0) & \rho_{01}(0) \\
            \rho_{10}(0) & \rho_{11}(0)
        \end{bmatrix}
\end{equation}
at the end of the laser pulse ($t=0$), implying that a certain amount of the ground-state population is transferred to $| \varphi_1 \rangle$
by the laser pulse. The ensuing field-free time evolution of the system for $t>0$ is described by the Lindblad master equation
\begin{align}
	\label{eq:Lindblad_detailed}
	\frac{\partial \rho_{00}}{\partial t} &= \gamma_\textrm{c} \rho_{11} \nonumber \\
	\frac{\partial \rho_{01}}{\partial t} &= \left( \textrm{i} \omega_\textrm{c} - \frac{\gamma_\textrm{c}}{2} \right) \rho_{01} \\	
	\frac{\partial \rho_{10}}{\partial t} &= \left( -\textrm{i} \omega_\textrm{c} - \frac{\gamma_\textrm{c}}{2} \right) \rho_{10} \nonumber \\	
	\frac{\partial \rho_{11}}{\partial t} &= - \gamma_\textrm{c} \rho_{11} \nonumber
\end{align}
in the basis of $| \varphi_0 \rangle$ and $| \varphi_1 \rangle$ (see also Eq. (98) in Ref. 58 of the manuscript).
The solution of Eq. \eqref{eq:Lindblad_detailed} reads
\begin{align}
	\rho_{00}(t) &= 1-\rho_{11}(0) \exp(-\gamma_\textrm{c} t) \nonumber \\
	\rho_{01}(t) &= \rho_{01}(0) \exp \left(-\frac{\gamma_\textrm{c}}{2} t \right) \exp(\textrm{i} \omega_\textrm{c} t) \\
	\rho_{10}(t) &= \rho_{10}(0) \exp \left(-\frac{\gamma_\textrm{c}}{2} t \right) \exp(-\textrm{i} \omega_\textrm{c} t) \nonumber \\
	\rho_{11}(t) &= \rho_{11}(0) \exp(-\gamma_\textrm{c} t) \nonumber.
\end{align}
In the basis of $| \varphi_0 \rangle$ and $| \varphi_1 \rangle$ the photon number operator $\hat{N}$ is represented by the matrix
\begin{equation}
	N = 
	\begin{bmatrix}
            0 & 0 \\
            0 & 1
        \end{bmatrix}
\end{equation}
with which the expectation value of $\hat{N}$ evaluates to
\begin{equation}
	N(t) = \textrm{Tr}[ \rho(t) N] = \rho_{11}(t) = \rho_{11}(0) \exp(-\gamma_\textrm{c} t).
  \label{eq:expval}
\end{equation}
Eq. \eqref{eq:expval} clearly shows that $N(t)$ decays exponentially with time in the adiabatic approximation
and the rate of the exponential decay is determined by the cavity decay rate ($\gamma_\textrm{c}$).

\clearpage

\section{Additional population and expectation value figures for $g=0.01 ~ \textrm{au}$ and $g=0.005 ~ \textrm{au}$}

\begin{figure}[hbt!]
\includegraphics[scale=0.8]{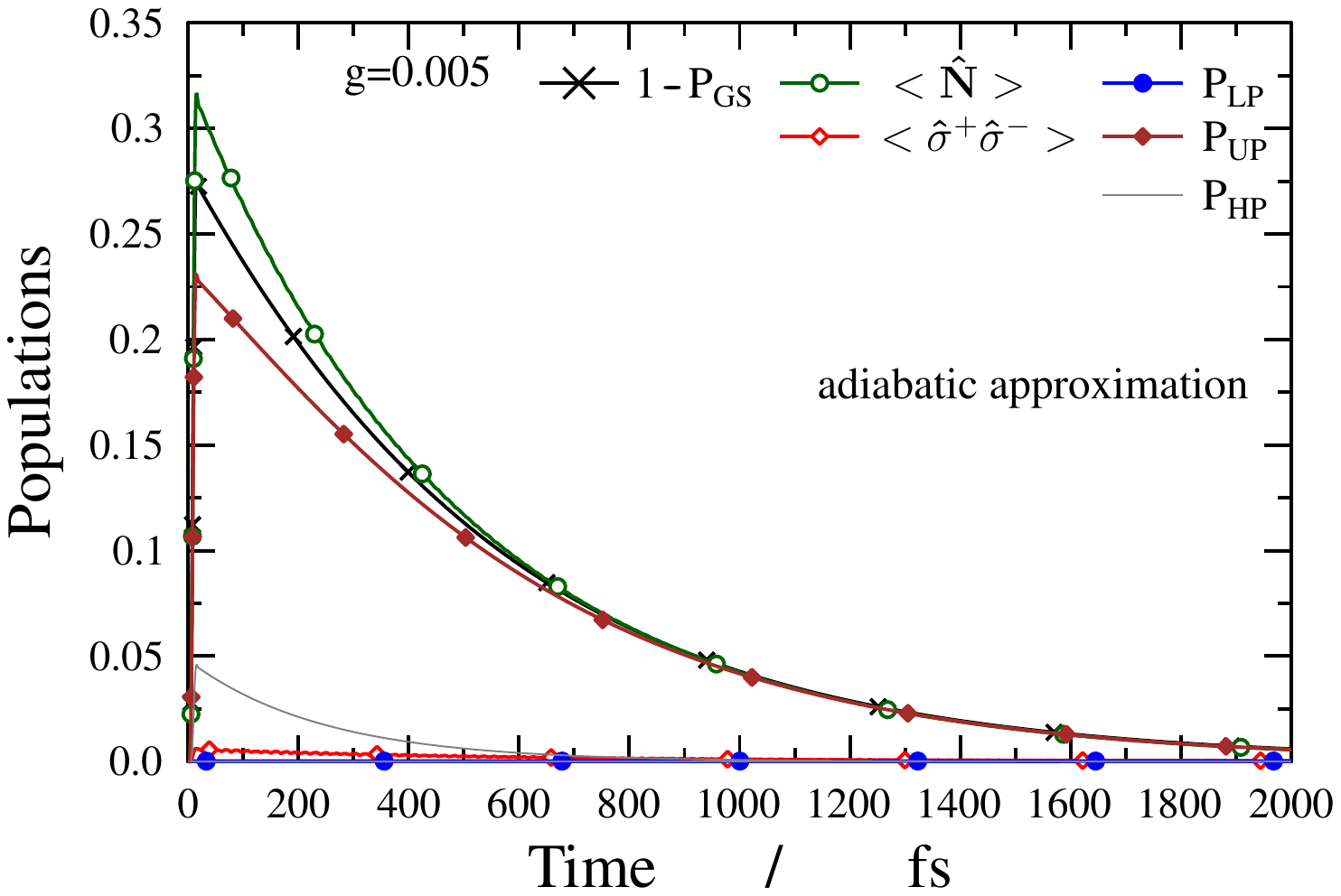}
\caption{\label{fig:pop_case2_ad}
Populations of polaritonic states (GS: ground-state (lowest) polariton, LP: lower polariton, UP: upper polariton, HP: higher-lying polaritons)
and expectation values of the operators $\hat{N}$ and $\hat{\sigma}^+ \hat{\sigma}^-$ in the adiabatic approximation
(cavity parameters: $\omega_\textrm{c} = 35744.8 ~ \textrm{cm}^{-1}$, $g=0.005 ~ \textrm{au}$ and $\gamma_\textrm{c} = 5 \cdot 10^{-5} ~ \textrm{au}$,
laser pulse parameters: $\omega = 36000 ~ \textrm{cm}^{-1}$, $T = 15 ~ \textrm{fs}$ and $E_0 = 3.77 \cdot 10^{-3} ~ \textrm{au}$).}
\end{figure}

Figure \ref{fig:pop_case2_ad} shows the population and expectation value
($\langle \hat{N} \rangle$ and $\langle \hat{\sigma}^+ \hat{\sigma}^- \rangle$) figures for $g=0.005 ~ \textrm{au}$
in the adiabatic approximation (see the figure caption for more information on the parameters).

\begin{figure}[hbt!]
\includegraphics[scale=0.8]{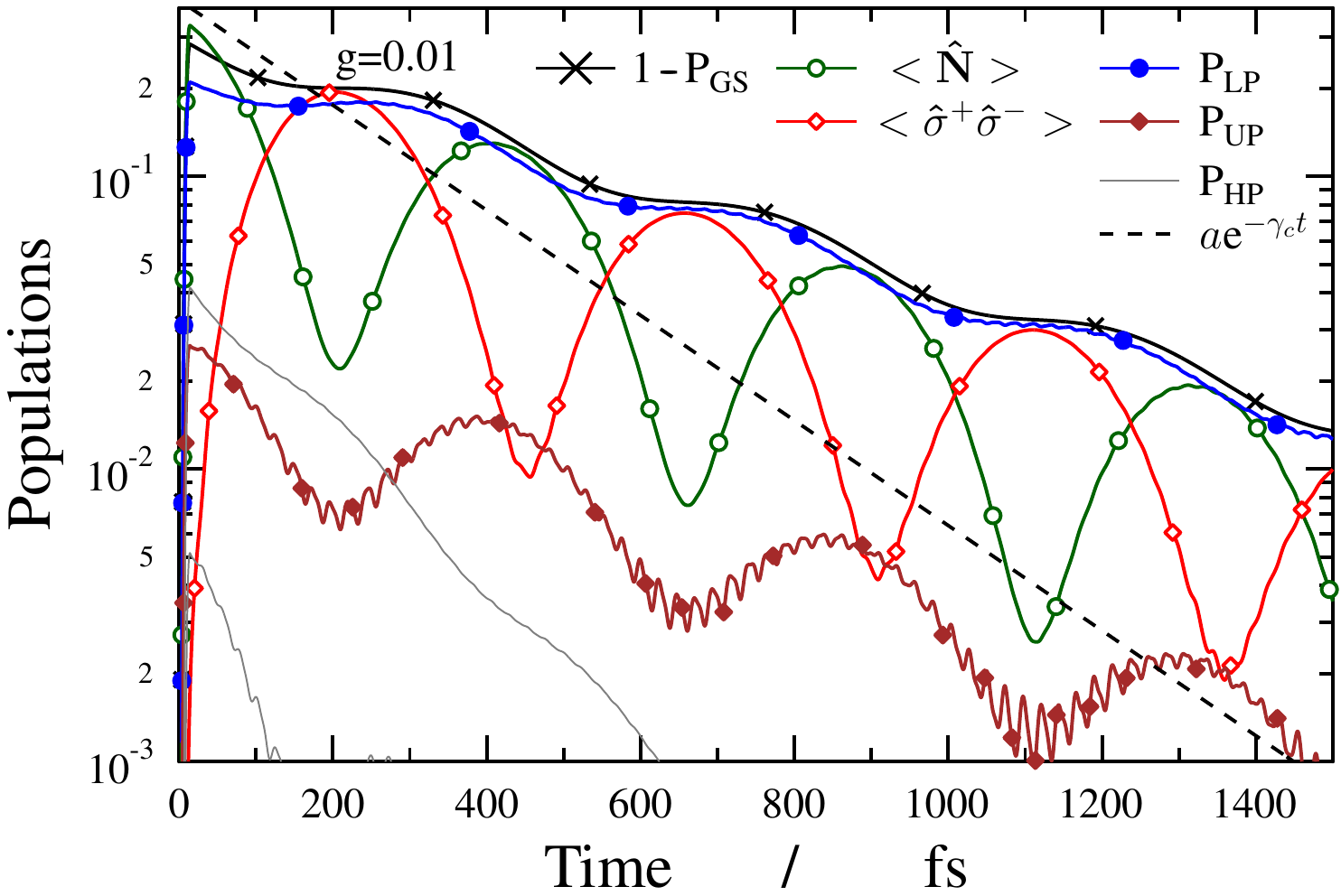}
\caption{\label{fig:poplog1}
Populations of polaritonic states (GS: ground-state (lowest) polariton, LP: lower polariton, UP: upper polariton, HP: higher-lying polaritons)
and expectation values of the operators $\hat{N}$ and $\hat{\sigma}^+ \hat{\sigma}^-$ on logarithmic scale
(cavity parameters: $\omega_\textrm{c} = 29957.2 ~ \textrm{cm}^{-1}$, $g=0.01 ~ \textrm{au}$ and $\gamma_\textrm{c} = 10^{-4} ~ \textrm{au}$,
laser pulse parameters: $\omega = 30000 ~ \textrm{cm}^{-1}$, $T = 15 ~ \textrm{fs}$ and $E_0 = 3.77 \cdot 10^{-3} ~ \textrm{au}$).}
\end{figure}

\begin{figure}[hbt!]
\includegraphics[scale=0.8]{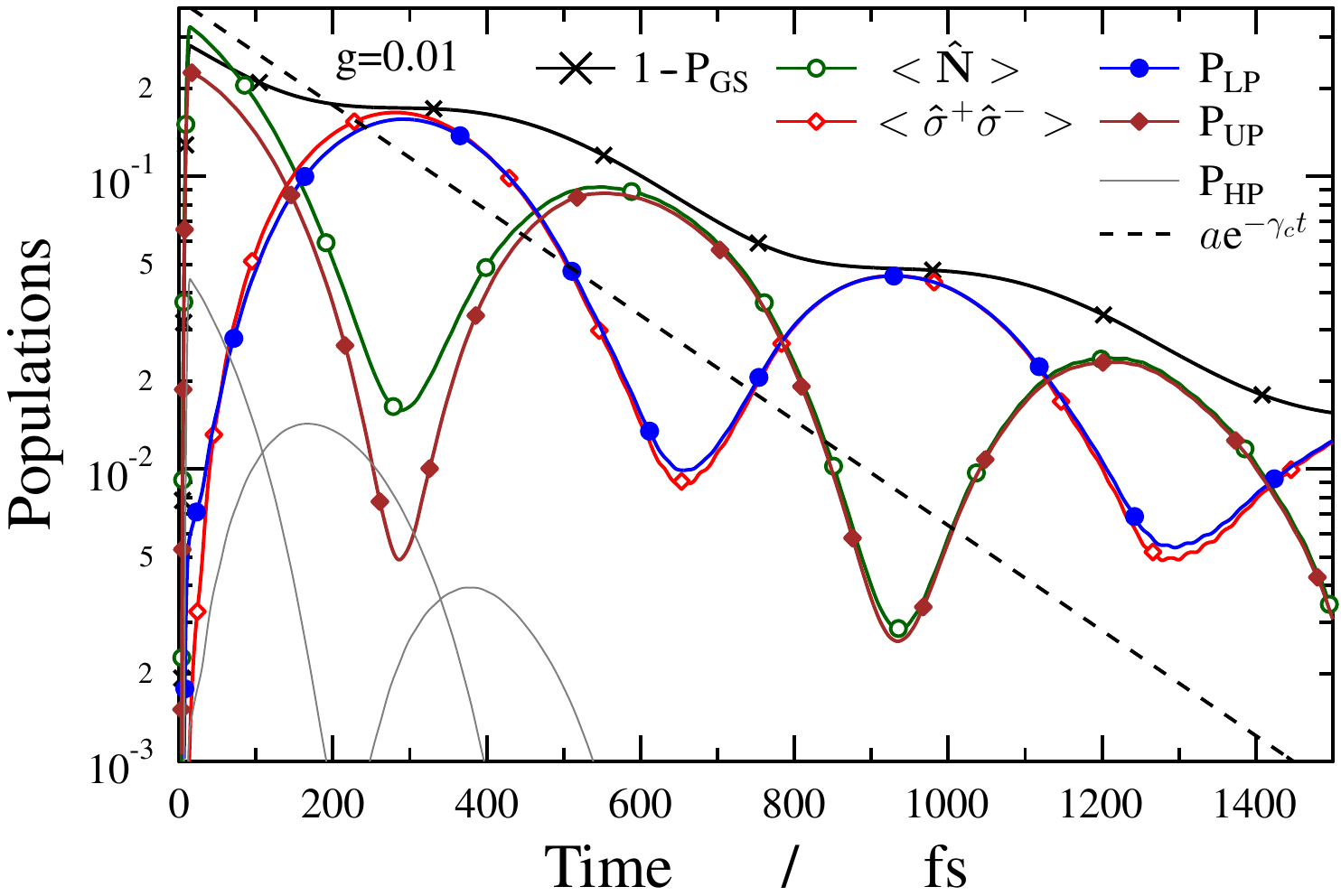}
\includegraphics[scale=0.8]{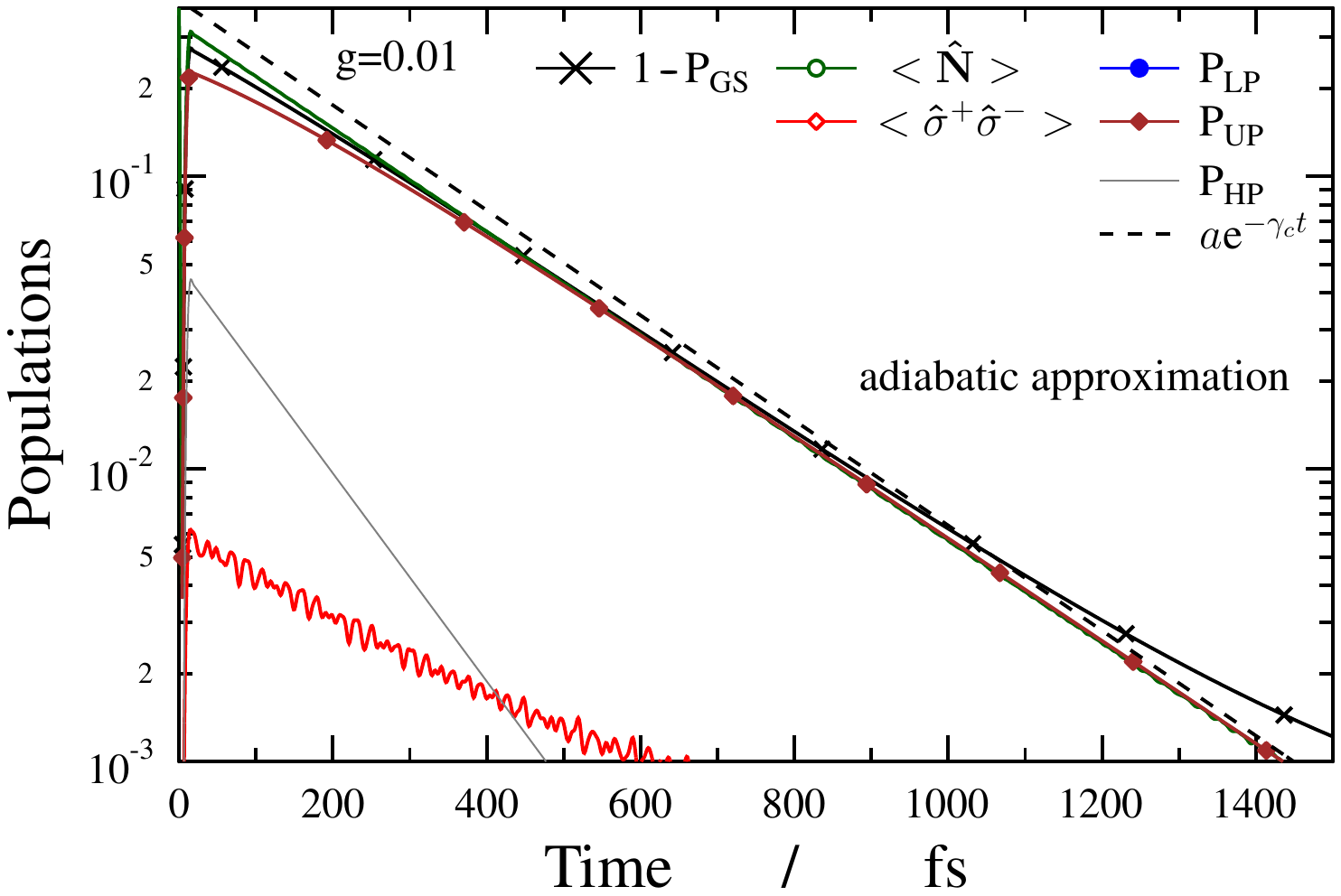}
\caption{\label{fig:poplog2}
Populations of polaritonic states (GS: ground-state (lowest) polariton, LP: lower polariton, UP: upper polariton, HP: higher-lying polaritons)
and expectation values of the operators $\hat{N}$ and $\hat{\sigma}^+ \hat{\sigma}^-$ on logarithmic scale
(cavity parameters: $\omega_\textrm{c} = 35744.8 ~ \textrm{cm}^{-1}$, $g=0.01 ~ \textrm{au}$ and $\gamma_\textrm{c} = 10^{-4} ~ \textrm{au}$,
laser pulse parameters: $\omega = 36000 ~ \textrm{cm}^{-1}$, $T = 15 ~ \textrm{fs}$ and $E_0 = 3.77 \cdot 10^{-3} ~ \textrm{au}$).
The upper and lower panels correspond to the exact case and the adiabatic approximation, respectively.}
\end{figure}

Figures \ref{fig:poplog1} and \ref{fig:poplog2} show the population and expectation value
($\langle \hat{N} \rangle$ and $\langle \hat{\sigma}^+ \hat{\sigma}^- \rangle$) figures for $g=0.01 ~ \textrm{au}$
on logarithmic scale (see the figure captions for more information on the parameters).

\clearpage
\section{Results for $g=0.1 ~ \textrm{au}$}

\begin{figure}[h]
\includegraphics[scale=0.65]{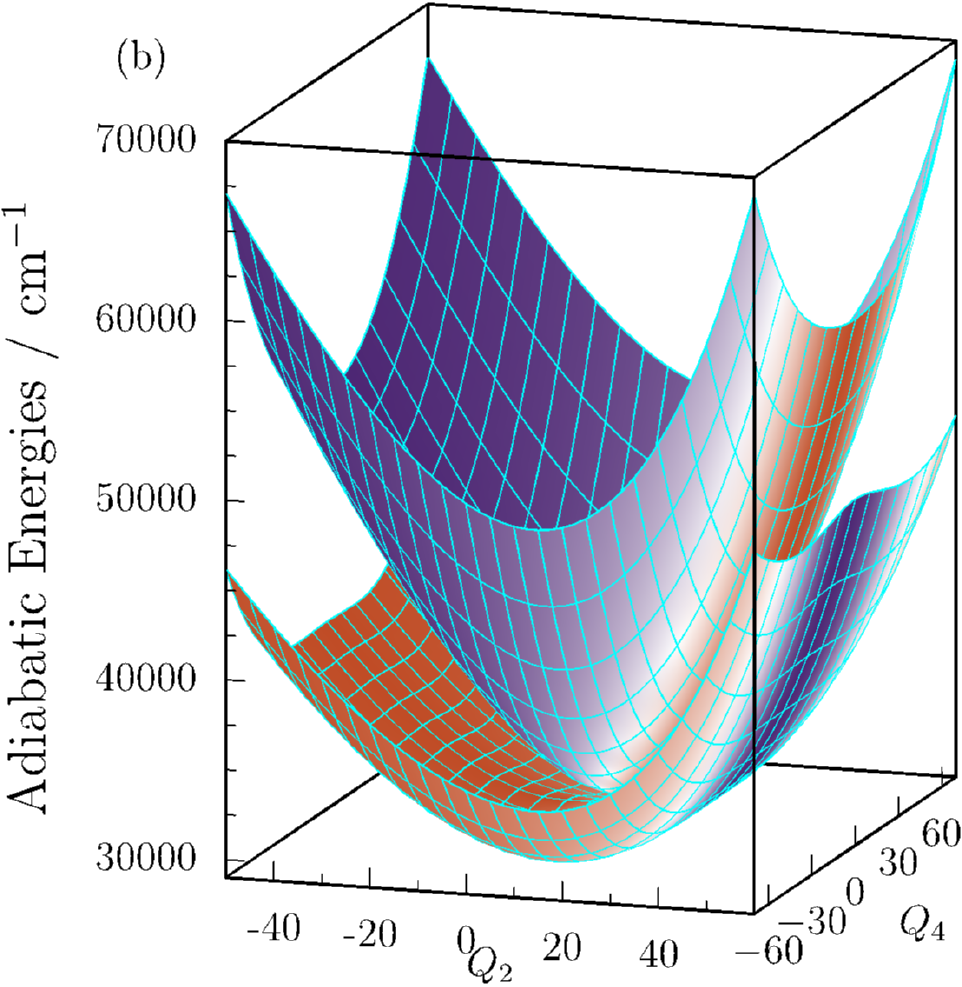}
\includegraphics[scale=0.65]{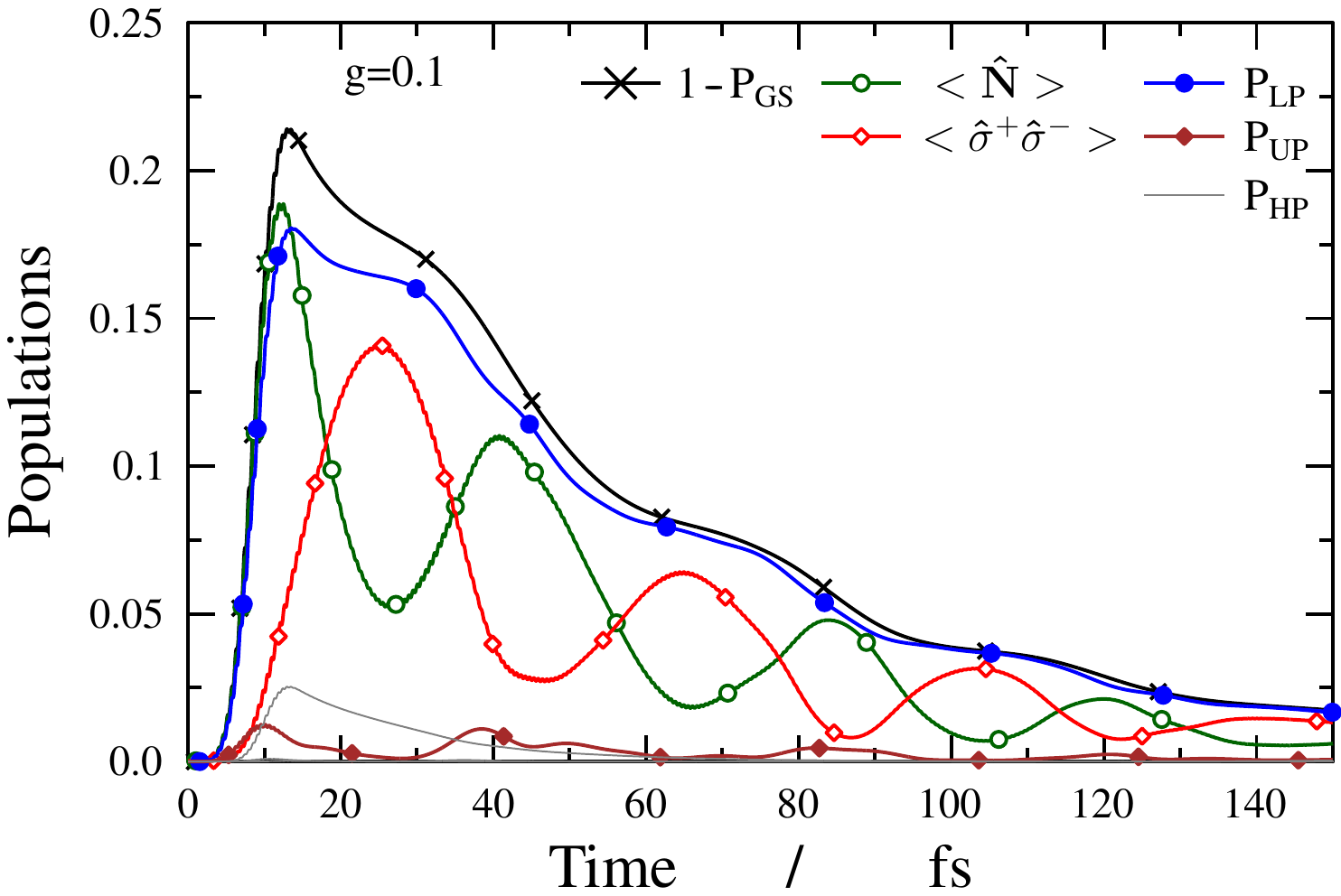}
\includegraphics[scale=0.65]{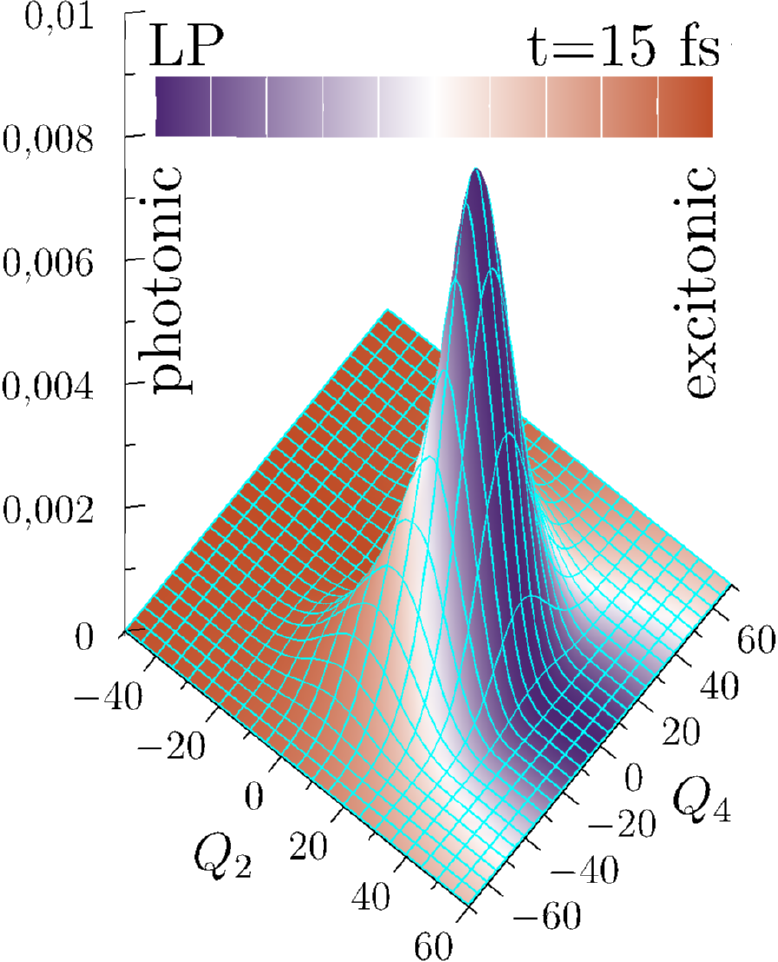}
\includegraphics[scale=0.65]{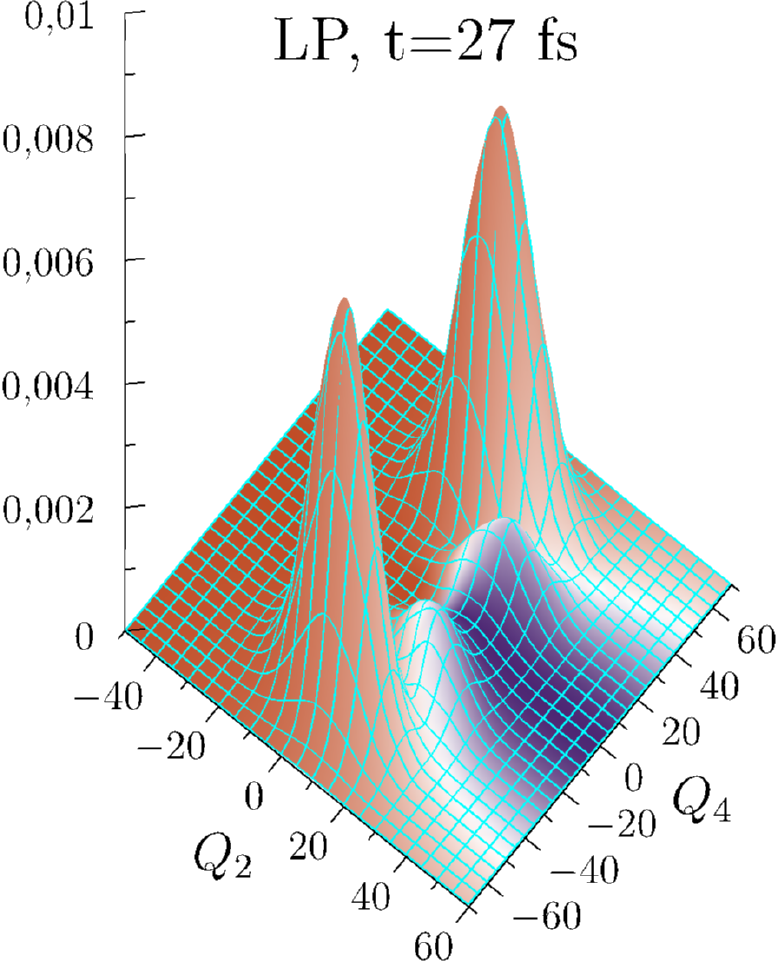}
\includegraphics[scale=0.65]{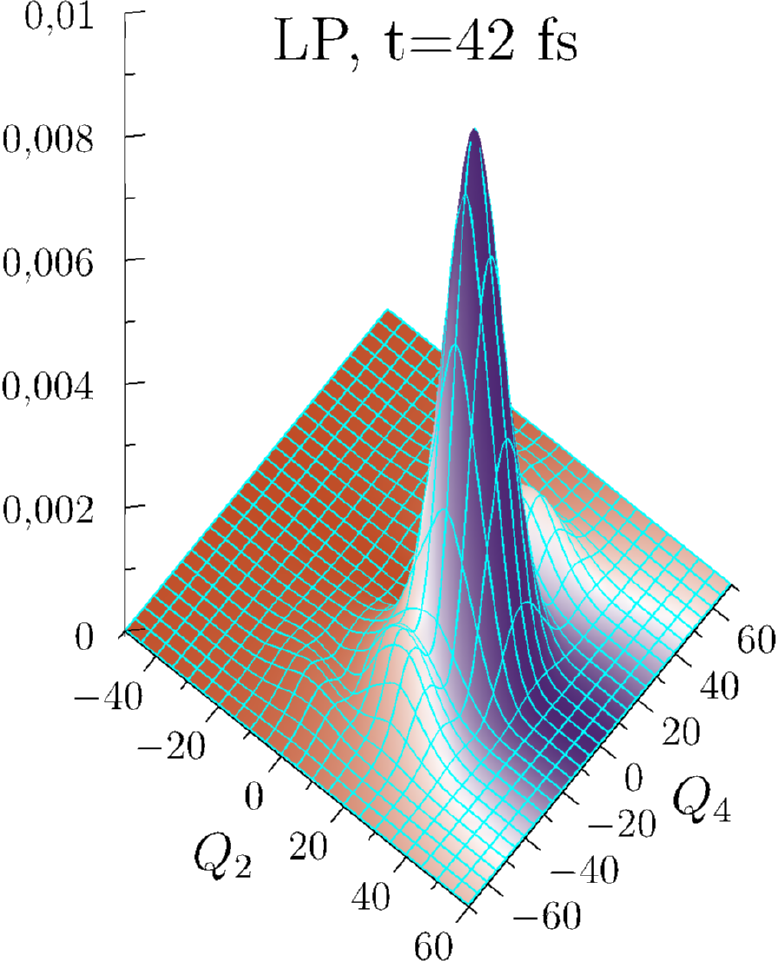}
\caption{\label{fig:case1_exact}
	Populations of polaritonic states (GS: ground-state (lowest) polariton, LP: lower polariton, UP: upper polariton, HP: higher-lying polaritons),
	expectation values of the operators $\hat{N}$ and $\hat{\sigma}^+ \hat{\sigma}^-$ and
        probability density figures for the LP surface (cavity parameters: $\omega_\textrm{c} = 29957.2 ~ \textrm{cm}^{-1}$, 
        $g=0.1 ~ \textrm{au}$ and $\gamma_\textrm{c} = 10^{-3} ~ \textrm{au}$,
        laser pulse parameters: $\omega = 30000 ~ \textrm{cm}^{-1}$, $T = 15 ~ \textrm{fs}$ and $E_0 = 3.77 \cdot 10^{-3} ~ \textrm{au}$).
        The orange and purple colors indicate excitonic and photonic characters of the polaritonic states, respectively.
        Time values are specified in each probability density panel.
}
\end{figure}

\begin{figure}[h]
\includegraphics[scale=0.6]{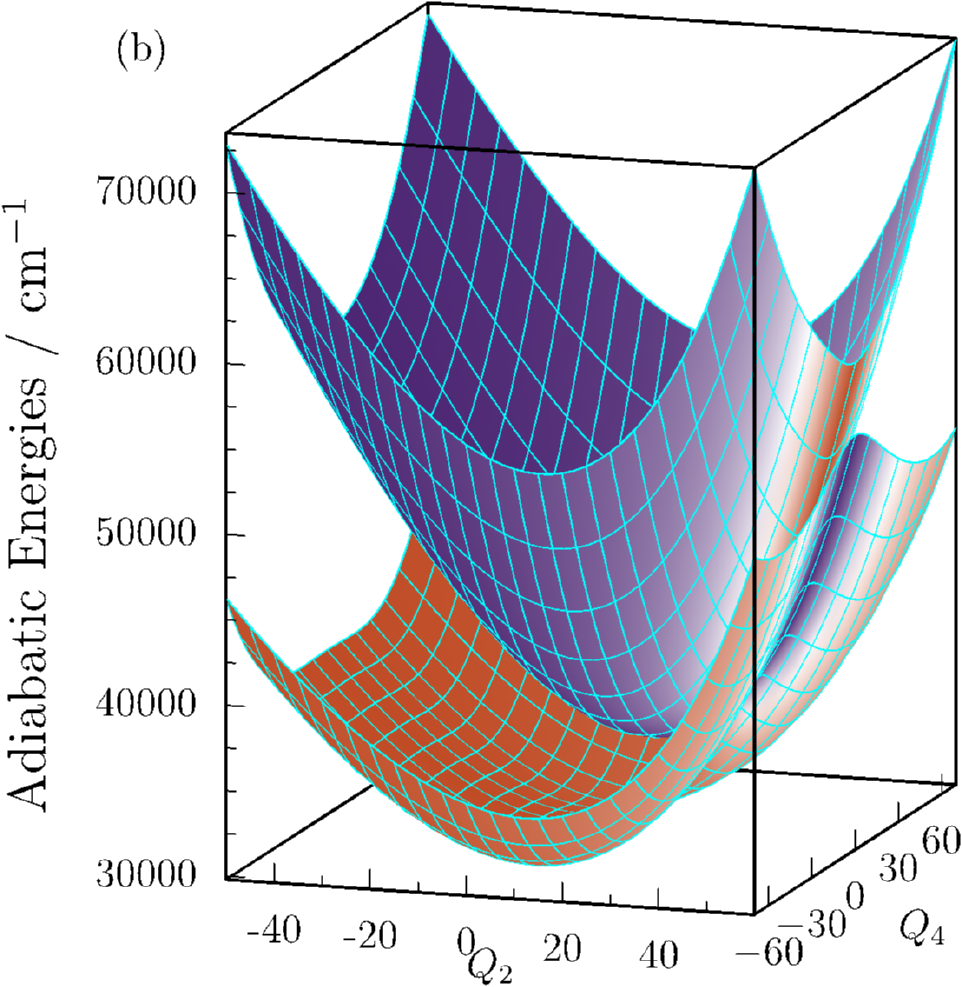}
\includegraphics[scale=0.6]{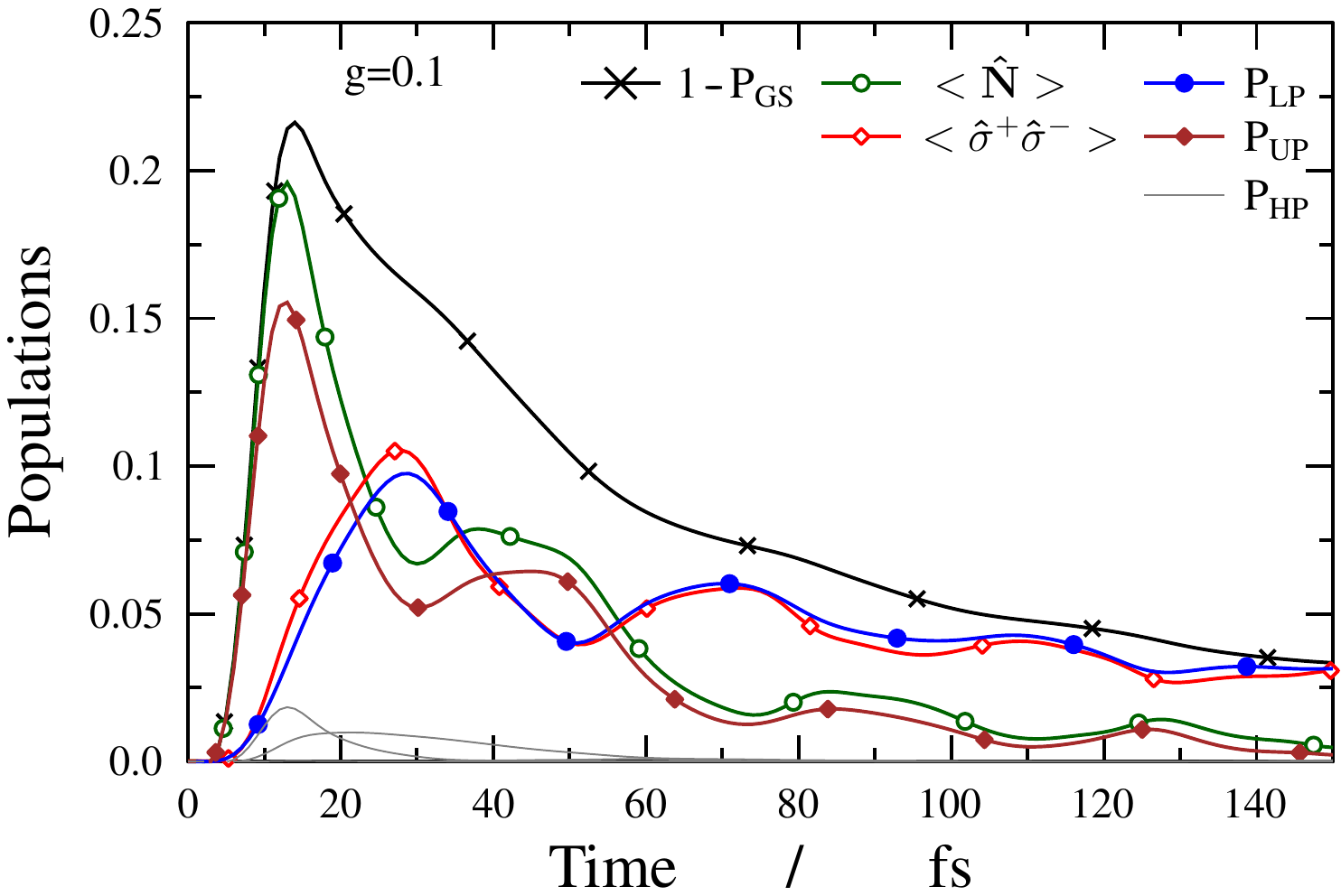}
\includegraphics[scale=0.65]{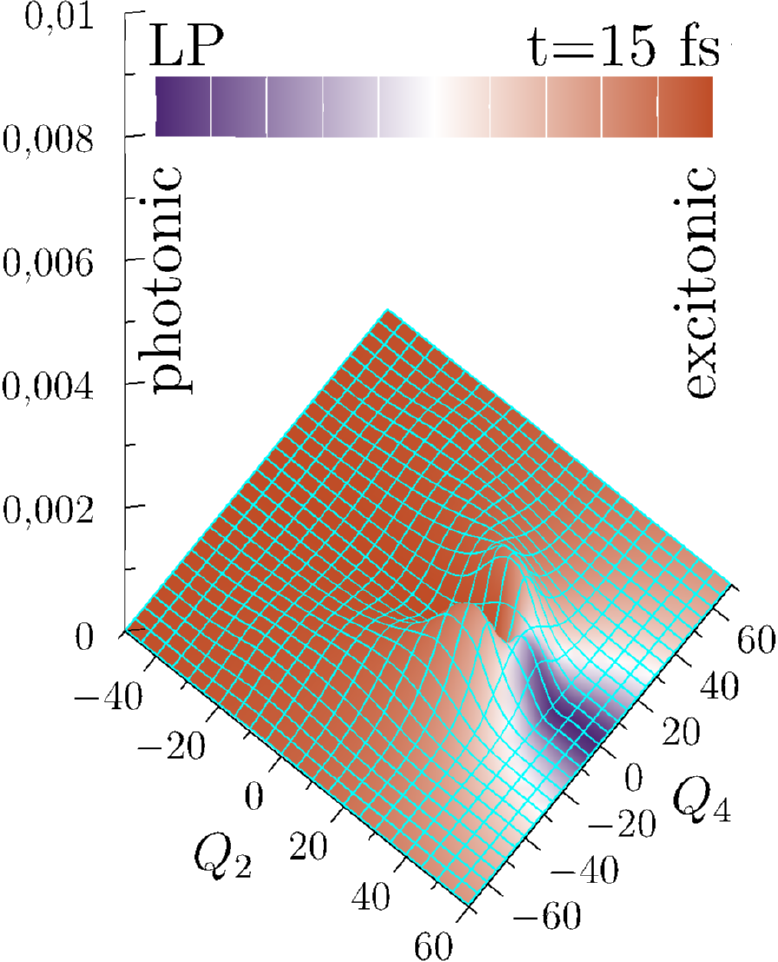}
\includegraphics[scale=0.65]{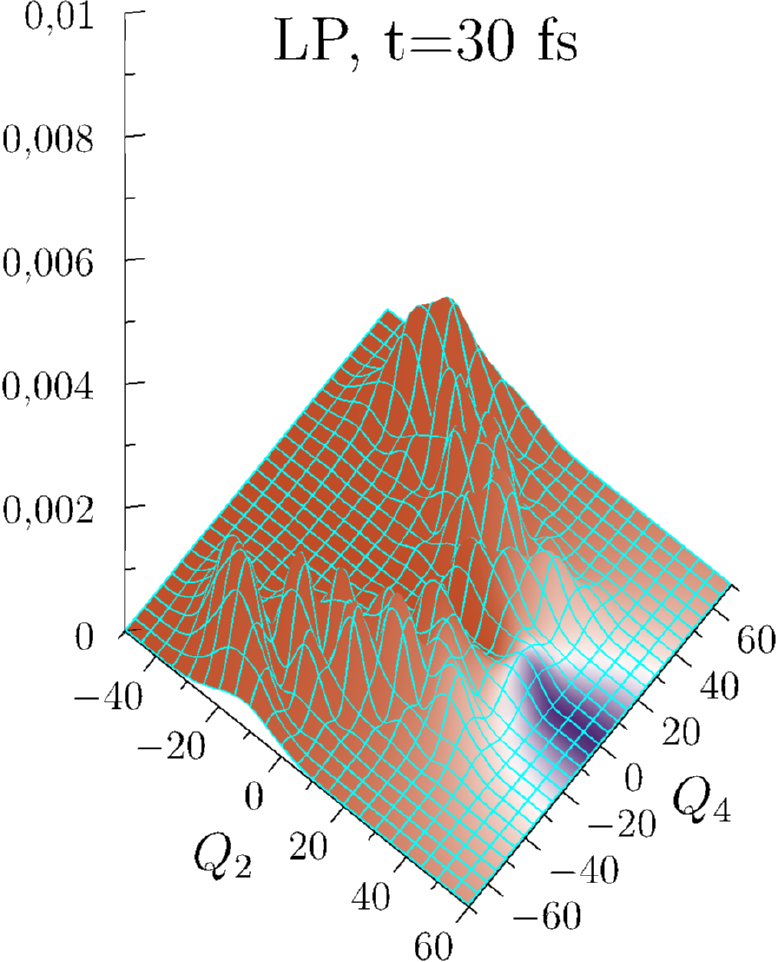}
\includegraphics[scale=0.65]{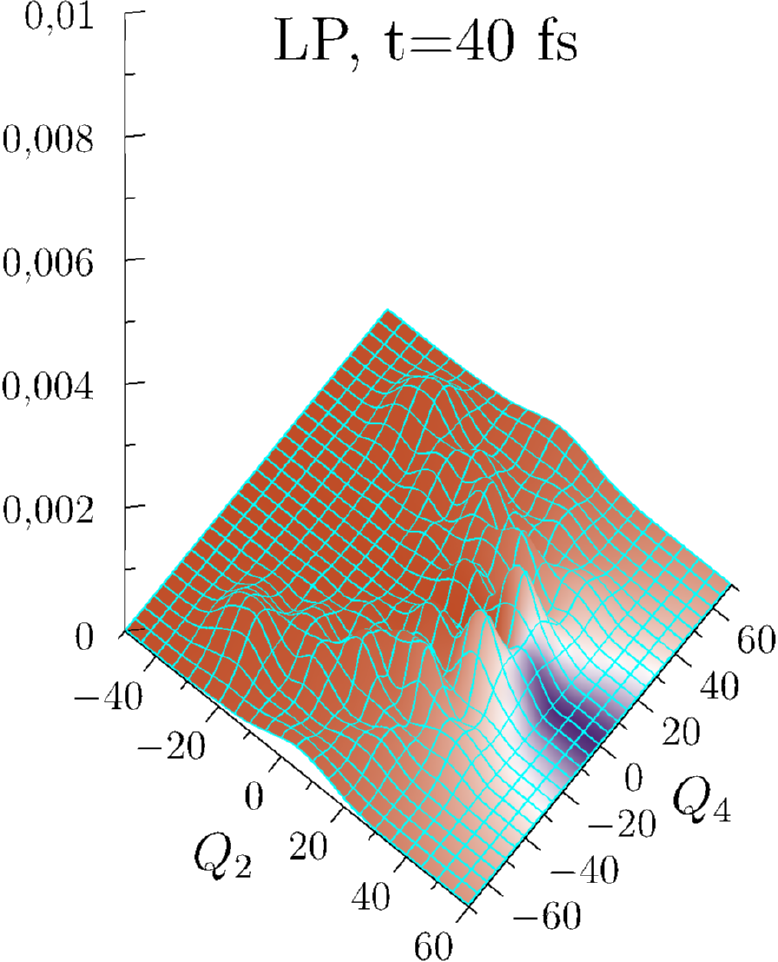}
\includegraphics[scale=0.65]{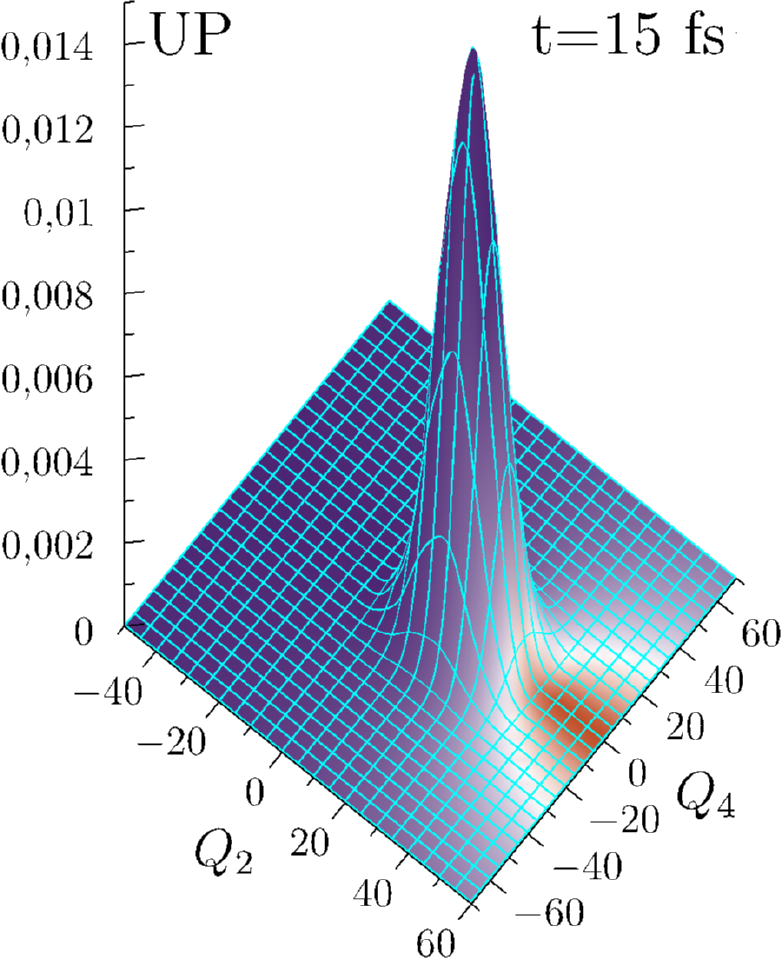}
\includegraphics[scale=0.65]{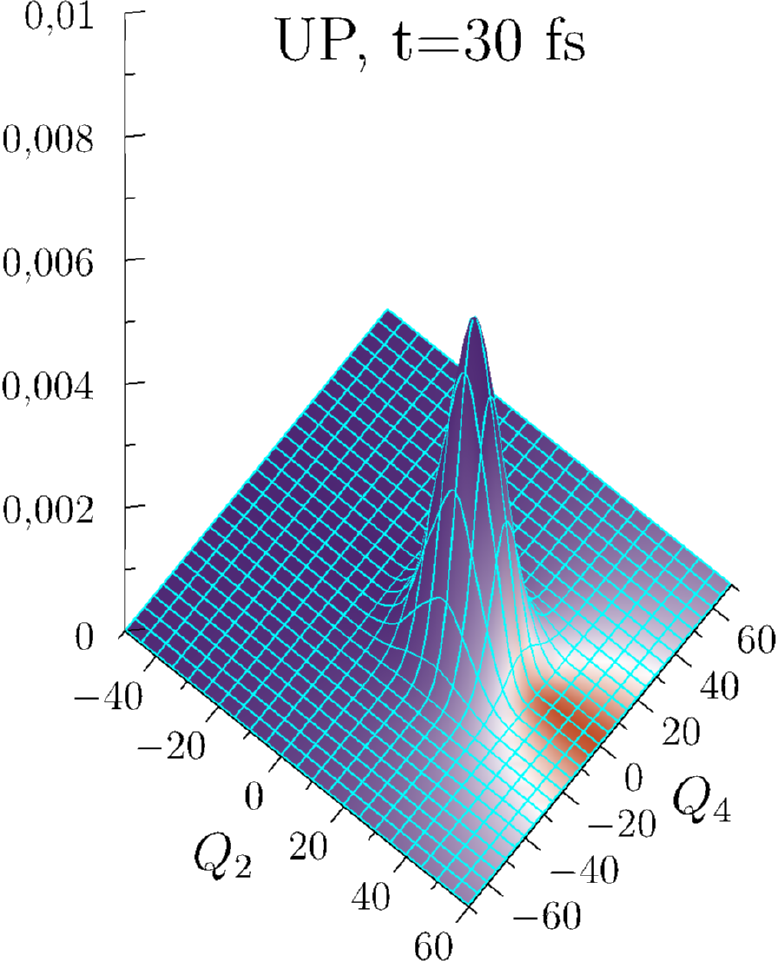}
\includegraphics[scale=0.65]{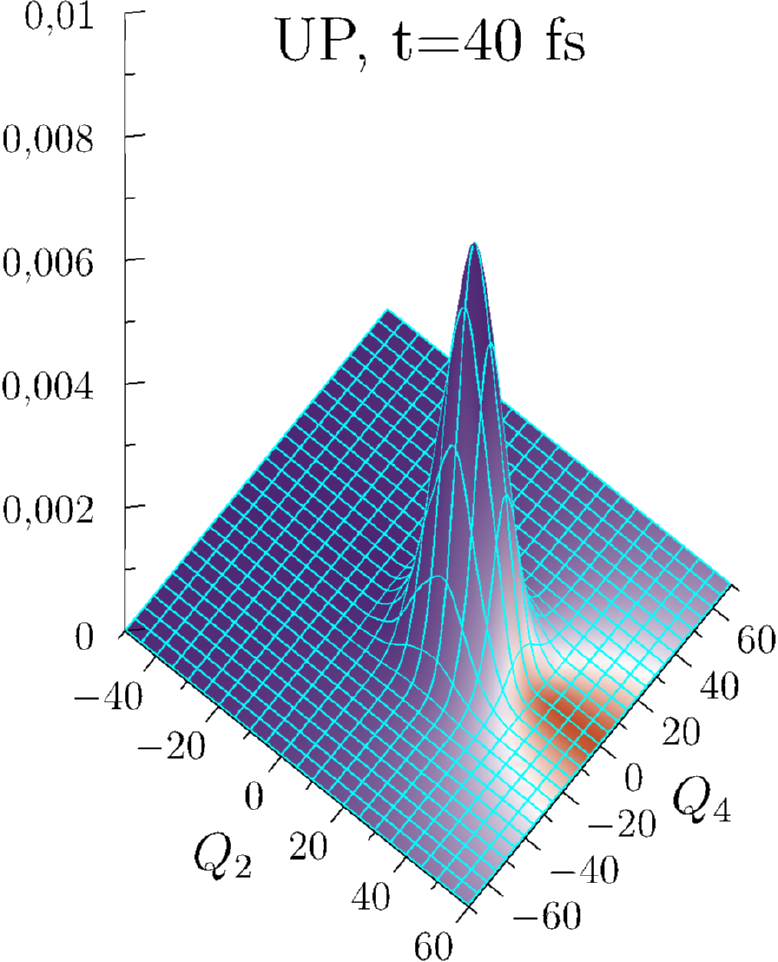}
\caption{\label{fig:case2_exact}        
        	Populations of polaritonic states (GS: ground-state (lowest) polariton, LP: lower polariton, UP: upper polariton, HP: higher-lying polaritons),
	expectation values of the operators $\hat{N}$ and $\hat{\sigma}^+ \hat{\sigma}^-$ and
        probability density figures for the LP and UP surfaces (cavity parameters: $\omega_\textrm{c} = 35744.8 ~ \textrm{cm}^{-1}$, 
        $g=0.1 ~ \textrm{au}$ and $\gamma_\textrm{c} = 10^{-3} ~ \textrm{au}$,
        laser pulse parameters: $\omega = 36000 ~ \textrm{cm}^{-1}$, $T = 15 ~ \textrm{fs}$ and $E_0 = 3.77 \cdot 10^{-3} ~ \textrm{au}$).
        The orange and purple colors indicate excitonic and photonic characters of the polaritonic states, respectively.
        Time values are specified in each probability density panel.
}
\end{figure}

Figures \ref{fig:case1_exact} and \ref{fig:case2_exact} show the population, expectation value
($\langle \hat{N} \rangle$ and $\langle \hat{\sigma}^+ \hat{\sigma}^- \rangle$) and probability density
figures for $g=0.1 ~ \textrm{au}$ (see the figure captions for more information on the parameters).
These results show essentially the same effects that are presented in the manuscript. Due to the (unphysically) high
coupling strength $g=0.1 ~ \textrm{au}$ the dynamics takes place on a shorter timescale than for
$g=0.01 ~ \textrm{au}$ and $g=0.005 ~ \textrm{au}$, which allows the use of
$\gamma_\textrm{c} = 10^{-3} ~ \textrm{au}$ (equivalent to a lifetime of $24.2 ~ \textrm{fs}$)
larger than the $\gamma_\textrm{c}$ values applied in the manuscript.

\clearpage
\section{Results for $g=0.002 ~ \textrm{au}$}

\begin{figure}[hbt!]
\includegraphics[scale=0.9]{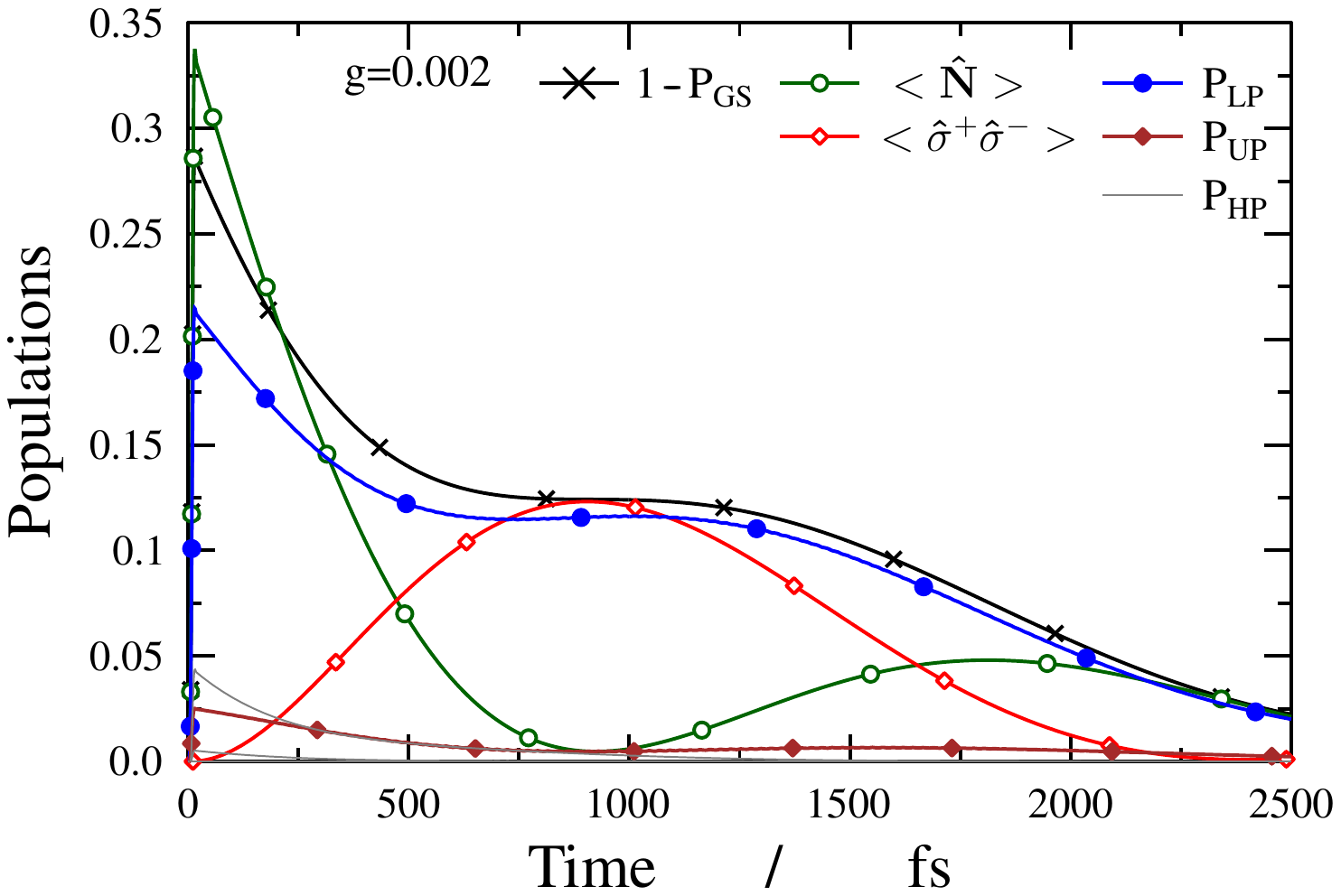}
\caption{\label{fig:pop0.002_case1}
Populations of polaritonic states (GS: ground-state (lowest) polariton, LP: lower polariton, UP: upper polariton, HP: higher-lying polaritons)
and expectation values of the operators $\hat{N}$ and $\hat{\sigma}^+ \hat{\sigma}^-$
(cavity parameters: $\omega_\textrm{c} = 29957.2 ~ \textrm{cm}^{-1}$, $g=0.002 ~ \textrm{au}$ and $\gamma_\textrm{c} = 5 \cdot 10^{-5} ~ \textrm{au}$,
laser pulse parameters: $\omega = 30000 ~ \textrm{cm}^{-1}$, $T = 15 ~ \textrm{fs}$ and $E_0 = 3.77 \cdot 10^{-3} ~ \textrm{au}$).}
\end{figure}

\begin{figure}[hbt!]
\includegraphics[scale=0.9]{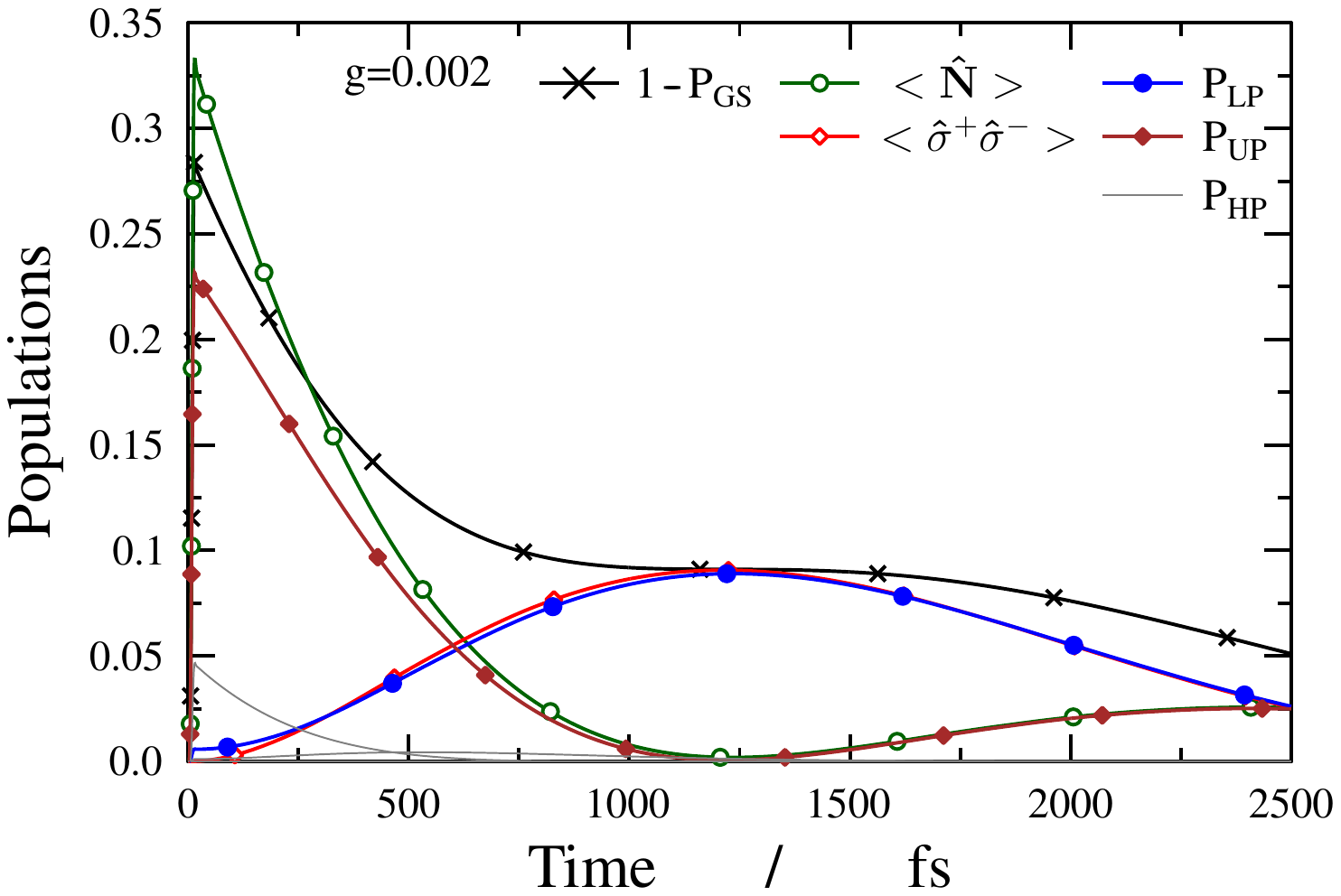}
\caption{\label{fig:pop0.002_case2}
Populations of polaritonic states (GS: ground-state (lowest) polariton, LP: lower polariton, UP: upper polariton, HP: higher-lying polaritons)
and expectation values of the operators $\hat{N}$ and $\hat{\sigma}^+ \hat{\sigma}^-$
(cavity parameters: $\omega_\textrm{c} = 35744.8 ~ \textrm{cm}^{-1}$, $g=0.002 ~ \textrm{au}$ and $\gamma_\textrm{c} = 5 \cdot 10^{-5} ~ \textrm{au}$,
laser pulse parameters: $\omega = 36000 ~ \textrm{cm}^{-1}$, $T = 15 ~ \textrm{fs}$ and $E_0 = 3.77 \cdot 10^{-3} ~ \textrm{au}$).}
\end{figure}

Figures \ref{fig:pop0.002_case1} and \ref{fig:pop0.002_case2} show the population and expectation value
($\langle \hat{N} \rangle$ and $\langle \hat{\sigma}^+ \hat{\sigma}^- \rangle$) figures for $g=0.002 ~ \textrm{au}$
(see the figure captions for more information on the parameters).
These results show essentially the same effects that are presented in the manuscript. Due to our choice of the
coupling strength $g=0.002 ~ \textrm{au}$ the dynamics takes place on a longer timescale than for
$g=0.01 ~ \textrm{au}$ and $g=0.005 ~ \textrm{au}$. As a consequence, oscillations in the emission signal
are smoothened out to a larger extent than for the $g=0.01 ~ \textrm{au}$ and $g=0.005 ~ \textrm{au}$ results.

\end{document}